\documentclass[aps,twocolumn,showpacs,preprintnumbers,amsmath,amssymb,nofootinbib,superscriptaddress,showkeys]{revtex4}

\usepackage{epsfig}
\usepackage{graphicx}
\usepackage{graphicx,amsmath,amssymb}

\begin{document}

\title{Nucleon-Nucleon interaction, charge symmetry breaking and
  renormalization} \author{A. Calle
  Cord\'on}\email{alvarocalle@ugr.es} \affiliation{ Departamento de
  F\'{\i}sica At\'omica, Molecular y Nuclear, Universidad de Granada,
  E-18071 Granada, Spain.}  \author{M. Pav\'on
  Valderrama}\email{m.pavon.valderrama@ific.uv.es}
\affiliation{Departamento de F\'{\i}sica Te\'orica and Instituto de
  F\'{\i}sica Corpuscular (IFIC), \\ Centro Mixto CSIC-Universidad de
  Valencia, \\ Institutos de Investigaci\'on de Paterna, Aptd. 22085,
  E-46071 Valencia, Spain} \author{E. Ruiz
  Arriola}\email{earriola@ugr.es} \affiliation{Departamento de
  F\'{\i}sica At\'omica, Molecular y Nuclear, Universidad de Granada,
  E-18071 Granada, Spain.}

\date{\today}

\begin{abstract} 
\rule{0ex}{3ex} We study the interplay between charge symmetry
breaking and renormalization in the $NN$ system for $s-$waves.  We
find a set of universality relations which disentangle explicitly the
known long distance dynamics from low energy parameters and extend
them to the Coulomb case.
We analyze within such an approach the One-Boson-Exchange potential
and the theoretical conditions which allow to relate the
proton-neutron, proton-proton and neutron-neutron scattering
observables without the introduction of extra new parameters and
providing good phenomenological success.
\end{abstract}


\maketitle



\section{Introduction} 

The understanding of Charge Dependence of strong interactions has been
a crucial issue in Nuclear Physics (for reviews see
e.g.~\cite{Wilkinson:1969,Miller:1990iz,Machleidt:2001rw,
  Miller:2006tv}).  In fact, the simplest place where this issue can
be studied is the nucleon-nucleon interaction. As it is well known,
isospin invariance is not an exact symmetry of strong interactions. As
a consequence nuclear forces have a small but net charge-dependent
component. By definition, \textit{charge independence} means
invariance under any rotation in isospin space. A violation of this
symmetry is referred to as charge dependence or charge independence
breaking (CIB) and it means in particular that, in the isospin $T = 1$
state, the proton-proton ($T_3 = +1$), neutron-proton ($T_3 = 0$), or
neutron-neutron ($T_3 = -1$) \textit{strong} interactions are
different. A particular case, known as charge symmetry breaking (CSB),
only considers the difference between proton-proton ($pp$) and
neutron-neutron ($nn$) interactions.  Further corrections are expected
when, in addition, Coulomb forces are added to the proton-proton
system ($pp(c)$).

But, what is the scale of charge symmetry breaking?. Actually, the
effects are important in the s-wave channel where an unnaturally large
scattering length, due to a virtual state~\footnote{That means a pole
  in the second Riemann sheet in the negative energy axis.} in that
partial wave, triggers a high short distance sensitivity. This, of
course amplifies effects related to variations in the short distance
parameters {\it precisely} in the region where the interaction and
hence the charge symmetry breaking effects may be less reliable. The
current understanding is that CIB, and in particular CSB, are due to a
mass difference between the up and down quarks and electromagnetic
interactions. On the hadronic level and in a One-Boson-Exchange (OBE)
based picture (such as e.g.~\cite{Machleidt:1987hj}), major causes of
CIB and CSB are effects explicitly related to
\begin{itemize}
\item Different proton and neutron masses. 
\item Electromagnetic effects (mainly Coulomb interaction). 
\item Mass splitting of isovector mesons $\pi$ and $\rho$ and different coupling
constants.
\item Mass splitting between different $\Delta$-isobar charge states.
\item Unknown short distance effects which are usually described by models. 
\end{itemize} 
Traditionally it is believed that the difference between the charge
and neutral pion mass in the One-Pion-Exchange (OPE) potential
accounts for a big part of CIB while the difference between the masses
of neutron and proton represents the most basic cause for CSB.
Pion mass differences were shown to account for a $80\%$ of the
$nn$-$pp$ scattering length difference~\cite{Cheung:1986wr}.
The nucleon mass splitting also generates a difference in the kinetic
energies.  Some recent OBE models only consider the differences coming
from nucleon mass splitting and kinematical
effects~\cite{Stoks:1994wp,Wiringa:1994wb}. However, these effects can
only explain about a $15\%$ of the empirical CSB. As a consequence
some models leave CSB unexplained~\cite{Stoks:1994wp} while others
simply introduce a term {\it ad hoc} to explain the remaining
contribution~\cite{Wiringa:1994wb}. In Ref.~\cite{Li:1998ya}
$2\pi$-exchange contributions, $\pi\rho$ diagrams and other
multi-meson exchanges including the $\Delta$-isobar as intermediate
states were considered to explain the empirical CSB value
accurately. In Ref.~\cite{Li:1998xh} $2\pi$-exchange contributions
with $\Delta$ were found to be noticeable to explain the empirical CIB
value being $3\pi$- and $4\pi$-exchanges negligible. The difficulties
arising in multi-meson exchange diagrams, in particular the energy
dependence that they create, were avoided in the Bonn
potential~\cite{Machleidt:1987hj} by introducing two effective
scalar-isoscalar $\sigma$-mesons simulating $2\pi + \pi\rho$
exchanges. In the highly successful CD-Bonn
potential~\cite{Machleidt:2000ge} CSB was included at the simplest
one-boson-exchange diagrams with the same philosophy as its
predecessor~\cite{Machleidt:1987hj}. 

Many authors have also proposed the $\rho-\omega$ mixing as a key
ingredient to understand CSB~\cite{Machleidt:2000vh,Biswas:2008ye}. In
Ref.~\cite{Machleidt:2000vh} the $\rho-\omega$ mixing is identified as
the major source of CSB while proton and nucleon mass differences are
identified to produce a minor effect. It should be noted that such a
calculation is hampered by the fact that the $g_{\omega NN}$ coupling
constant occurring in the CS part is about $40\%$ larger than expected
from $SU(3)$ ($g_{\omega NN} = 3 g_{\rho NN} \sim 9$) and also from
the actual value taken for the CSB potential. Fixed-s Dispersion
relations~\cite{Hamilton:1983ri} yielded $g_{\omega NN} = 5.7 \pm 2.0
$ and Vector Meson Dominance $\omega \to e^+e^-$ decays prefer
$g_{\omega NN} \sim 10$. The $\eta-\pi^0$ has been shown to be of some
relevance as well~\cite{Piekarewicz:1993ad}.

The high sensitivity has been a major motivation to pursue
experimental determinations of the neutron-neutron scattering length
by indirect methods (for a review see e.g. Ref.~\cite{Howell:2008dt}
and references therein). A recent measurement of the nn scattering
length using the pi-d capture reaction, yields (see also
\cite{Gardestig:2009ya} for a review ) $a_{nn}= -18.69(4)\,  {\rm fm}$ when
corrected for magnetic interactions.  The recent CSB analysis of the
reaction $dd \to \pi^0 \alpha$~\cite{Fonseca:2009qs} uses also the
large $g_{\omega NN}$ constant.

The purpose of the present work is to approach the problem from a
renormalization point of view. While we consider long distance physics
to be {\it known} and describable by non-relativistic potentials we
use a physical low energy parameter such as the scattering length to
encode the unknown short distance physics. For the Charge Independent
case it was shown how a natural $SU(3)$ value of $g_{\omega NN}$ could
be confortably taken if a certain renormalization condition was
imposed~\cite{CalleCordon:2008eu,Cordon:2009pj}.  As we will discuss in
much detail this poses a problem of finiteness in physical observables
when connecting different channels such as $np$, $nn$ and $pp$ (strong
or Coulomb).
We propose a short distance renormalization condition featuring charge
independence which guarantees finiteness although an ambiguity arises.
However, a natural choice of the renormalization condition works quite
well when compared to measured or recommended values.  While the
traditional point of view outlined above tried to compute the
scattering lengths, the Effective Field Theory (EFT) approach assumes
that these scattering lengths are completely
unrelated~\cite{Kong:1998sx,Kong:1999sf,Walzl:2000cx,
  Gegelia:2003ta,Epelbaum:2005fd,Ando:2007fh}.  We pursue here the
possible connection between them from a new perspective which actually
is in-between, combining both points of view.  We assume one
scattering length to be known and exploit the concept of short
distance insensitivity to determine all other scattering lengths and
phase shifts from the requirement of finiteness of the scattering
amplitude.

The paper is organized as follows. In Section~\ref{sec:OBE} we
motivate the use of renormalization without assuming previous
knowledge from the reader as a useful preparatory material for further
developments. In Section~\ref{sec:Coulomb} we extend the results to
the interesting case of long range Coulomb interactions as they appear
in $pp$ scattering. CSB interactions are studied in all separate
$np$,$nn$,$pp$ and $pp(c)$ cases in Section~\ref{sec:CSB} where
insightfull universal relations are found out. In
Section~\ref{sec:SDC} we propose a short distance connection where all
the channels are correlated from the requirement of finiteness. A
further interesting application has to do with the determination in
Appendix~\ref{sec:ppF} of the Gamow-Teller matrix element needed in
the pp fusion process from the np scattering length. Finally, in
Section~\ref{sec:concl} we come to the conclusions.

\section{The standard vs. the renormalization approach}
\label{sec:OBE}

\subsection{The OBE potential}

In this section we briefly review the main ideas behind
renormalization in coordinate space for the OBE potentials (for a more
detailed account see e.g. Ref.~\cite{Cordon:2009pj}) since they are
focal in what follows.
To provide a comprehensive perspective we compare it with the more
traditional viewpoint of regulating the singular meson-exchange
potentials by means of the introduction of short-distance form
factors.
The crucial distinction lies in the sensitivity to short-distance
details: from the renormalization point of view we expect complete
insensitivity to these details.
On the contrary, a regularization procedure only guarantees
the finiteness of the results.
For definiteness, let us analyze as an illustrative example
the phenomenologically successful $^1S_0$ one boson exchange (OBE)
potential~\cite{Machleidt:2000ge,Machleidt:1987hj}
\begin{eqnarray}
V (r) &=& -\frac{ g_{\pi NN}^2 m_{\pi}^2} {16 \pi M_N^2}
\frac{e^{-m_{\pi} r}} {r} - \frac{ g_{\sigma NN} ^2}{4 \pi}\frac{e^{-
m_{\sigma} r}} {r}  
\nonumber \\ 
&+& \frac{g_{\omega NN}^2}{4 \pi}\frac{e^{-m_{\omega}
r}}{r} -
\frac{f_{\rho NN}^2 m_\rho^2}{8 \pi M_N^2 }\frac{e^{-m_{\rho}
r}}{r} \, ,
\label{eq:potential-OBE}
\end{eqnarray}
where $g_{\sigma NN}$ is a scalar type coupling, $g_{\pi NN}$ a
pseudo-scalar  coupling, $g_{\omega NN}$ is a vector
coupling and $f_{\rho NN}$ is a tensor derivative coupling (see
\cite{Machleidt:1987hj} for notation).
We neglect for simplicity nucleon mass effects and a tiny $\eta$
contribution.
We take $m_\pi=138 {\rm MeV}$, $M_N=939 {\rm MeV}$, $m_\rho=770 {\rm MeV}$, 
$m_\omega=783 {\rm MeV}$ and $g_{\pi NN}=13.1$ which seem firmly established.
The OBE potential, Eq.~(\ref{eq:potential-OBE}) corresponds
to a long distance expansion of the potential.
On the other hand, NN scattering in the elastic region below pion production
threshold involves CM momenta $p < p_{\rm max} = 400 {\rm MeV}$.
Given the fact that 
$1/m_\omega = 0.25\,{\rm fm} \ll 1/p_{\rm max}= 0.5\,{\rm fm}$
we expect heavier mesons to be irrelevant, and $\rho $ and
$\omega$ themselves to be of marginal important.
This naive expectation is, however, not fulfilled in the traditional
approach~\cite{Machleidt:2000ge,Machleidt:1987hj}.

In the following we will make the approximation $m_{\rho} = m_{\omega}$,
specially when making fits of the coupling parameters to the $^1S_0$ phases.
Under the previous approximation is convenient to define
\begin{eqnarray}
g_{\omega NN}^{*} =
\sqrt{g_{\omega NN}^2 - \frac{f_{\rho NN}^2 m_\rho^2}{2 M_N^2 }} \, ,
\end{eqnarray}
in such a way that the combined $\omega-\rho$ potential reads
\begin{eqnarray}
\frac{g_{\omega NN}^2}{4 \pi}\frac{e^{-m_{\omega}r}}{r} -
\frac{f_{\rho NN}^2 m_\rho^2}{8 \pi M_N^2 }\frac{e^{-m_{\rho}
r}}{r} \simeq \frac{{g_{\omega NN}^{*}}^2}{4 \pi}\frac{e^{-m_{\omega}r}}{r}
\, .
\end{eqnarray}
The previous simplification is useful since it avoids correlations
between $g_{\omega NN}$ and $f_{\rho NN}$ in the $^1S_0$ channel. For
SU(3) values of $g_{\omega NN} \sim 9$ or VMD $ \omega \to e^+e^- $ e.m. 
decays $g_{\omega NN} \sim 10.5$ and typical $\rho$-tensor
values $f_{\rho NN} \sim 14-18 $ one has $g_{\omega NN}^* \sim 0-7$. 

\subsection{Regular solution}

In the traditional approach~\cite{Machleidt:1987hj,Machleidt:2000ge}
the problem is essentially handled by solving the reduced
Schr\"odinger equation, which for the $s-$wave case reads
\begin{eqnarray} 
-u_k''(r) + M_N\,V(r) u_k(r) = k^2 u_k(r) \, ,
\label{eq:schroe-s}
\end{eqnarray}
with $k=\sqrt{M_N E}$ the center-of-mass momentum, $M_N$ the nucleon
mass, and $V(r)$ the boson-exchange potential of
Eq.~(\ref{eq:potential-OBE}).
The Schr\"odinger equation is a second order differential equation and
it has two linearly independent solutions.
The physical solution is usually determined by the regularity condition
at the origin, i.e.
\begin{eqnarray}
u_k(0) = 0 \, .
\end{eqnarray}
This boundary condition for the Schr\"odinger equation implicitly
assumes that we are taking the potential seriously all the way down to
the origin~\footnote{Of course, in a more conventional setup strong
  form factors accounting for the finite nucleon size should be
  included. We will argue below that they play a marginal role in the
  discussion of CSB.}.

The asymptotic behaviour of the reduced wave function for $r \gg 1/m_\pi$
is given by
\begin{eqnarray} 
u_k (r) \to \frac{\sin ( k r + \delta_0 (k))}{\sin \delta_0 (k)}
\label{eq:d0} \, ,
\end{eqnarray} 
where $\delta_0(k)$ is the s-wave phase shift.
For the potential described by Eq.~(\ref{eq:potential-OBE}),
the phase shift is an analytic function of $k$ with branch cuts
located at $k = \pm i m_\pi /2$, $\pm i m_{\sigma} / 2$, etc.
This means in particular that for momenta below the first branch cut,
$|k| \le m_\pi /2$, we can expand the phase shift by means of
the effective range expansion~\cite{PhysRev.76.38}
\begin{eqnarray}
k\,\cot \delta_0 (k) = 
- \frac{1}{\alpha_0} + \frac{1}{2} r_0 k^2 + v_2 k^4 + \dots \, ,
\label{eq:ere}
\end{eqnarray} 
where $\alpha_0$ is the scattering length, $r_0$ the effective range and
$v_2$ the shape parameter.

The effective range parameters can be related with the expansion of
the wave function in terms of $k^2$, i.e.
\begin{eqnarray}
\label{eq:uk_exp}
u_k (r)= u_0 (r) + k^2 u_2 (r) + \dots \, ,
\end{eqnarray}
where $u_0$ and $u_2$ obey the following equations
\begin{eqnarray} 
-u_0''(r) + M_N\,V(r) u_0(r) &=& 0 \, , \label{eq:u0} \\  
-u_2 '' (r) + M_N\,V (r) u_2 (r) &=& u_0 (r) \, , \label{eq:u2}
\end{eqnarray}
subjected to regular boundary conditions, $u_0(0) = u_2(0) = 0$,
and asymptotically normalized to
\begin{eqnarray}
u_0 (r) &\to& 1- r /\alpha_0  \, , \\ 
u_2 (r) &\to& \left(r^3
-3 \alpha_0 r^2 + 3 \alpha_0 r_0 r \right)/(6 \alpha_0) \, ,
\end{eqnarray}
for $r \gg 1/m_{\pi}$.
With this normalization, the effective range $r_0$ is computed from
the standard formula
\begin{eqnarray}
r_0 &=& 2\int_0^\infty dr \left[  \left(1- r
/\alpha_0\right)^2 - u_0 (r)^2\right] \label{eq:r0} \, .
\end{eqnarray} 
In the traditional approach~\cite{Machleidt:1987hj,Machleidt:2000ge}
{\it everything} is obtained from the potential,
which is assumed to be valid for $0 \le r < \infty$. 
In practice, strong form factors are included mimicking the finite nucleon size
and reducing the short distance repulsion of the potential,
but the regular boundary condition is always kept.
One of the problems with this point of view has to do
with the fact that the $^1S_0$ scattering length is unnaturally large
$\alpha_0 = -23.74(2) {\rm fm}$, while the effective range is natural,
$r_0= 2.77(4) {\rm fm}$ (approximately twice the pion Compton wave length,
$\sim 2 / m_{\pi}$).
This has dramatic consequences regarding the short
distance sensitivity, as we will show below. 

\begin{centering}
\begin{table*}
\begin{tabular}{|c|c|c|c|c|c|c|c|}
\hline\hline
BC &
 $r_c ({\rm fm})$ & $m_\sigma ({\rm MeV})$ & $g_{\sigma NN}$ &
 $g_{\omega NN}^*$ & $\chi^2 /DOF $ & $\alpha_0 ({\rm fm})$ & $r_0 ({\rm fm})$
\\ \hline 
Regular solution-I &
0 & 498.2(7) & 9.488(11) & 7.94(2) & 0.480 & -23.737 & 2.678 \\
\hline
Regular solution-II &
 0  & 550.72(4) & 13.87(13) & 20.10 (24) & 0.674 & -23.738 & 2.679 \\
\hline
Renormalizing &
 0  &  490(17) &  8.7(6) &  0(5) & 0.289 & \text{input} &  2.672 \\
\hline\hline
\end{tabular}
\caption{\label{tab:table_fits_CSB} Fits to the $^1S_0$ phase shift of the
Nijmegen group~\cite{Stoks:1994wp} using the OBE potential with a
charge dependent OPE part. We take $m_{\pi^0}= 134.97\,{\rm MeV}$, $m_{\pi^+}=
139.57\,{\rm MeV}$, $g_A = 1.29$ and
 $f_\pi = 92.4\,{\rm MeV}$~\cite{deSwart:1997ep}. We neglect the CSB coming
from the $\rho$-meson and take $m_\rho=m_\omega=770\,{\rm MeV}$ fitting
$m_\sigma$, $g_{\sigma NN}$ and $g_{\omega NN}^*$. We use the value 
$\alpha_{np} = - 23.74\,{\rm MeV}$ as an input when renormalizing.}
\end{table*}
\end{centering}

A fit to the $np$ averaged data of Ref.~\cite{Stoks:1994wp} in the
$^1S_0$ channel yields two possible solutions, see
Table~\ref{tab:table_fits_CSB}. Thus, we have two {\it good
  incompatible fits}. A remarkable aspect is the fact that the vector
meson coupling constant is accurately well determined.  Actually, if
we assume that we have fitted the potential,
Eq.~(\ref{eq:potential-OBE}), to reproduce $\alpha_0$, a tiny change
in the potential $V \to V + \Delta V$ has a dramatic effect on
$\alpha_0$, since one obtains
\begin{eqnarray}
\Delta \alpha_0 = \alpha_0^2 M_N \int_0^\infty \Delta V(r) u_0(r)^2 dr  \, ,
\label{eq:delta-alpha0} 
\end{eqnarray} 
a quadratic effect in the large $\alpha_0$.  As a result, potential
parameters must be fine tuned, as can be deduced from the previous
fits. Thus, despite the undeniable success in fitting the data, this
sensitivity to short distances looks counterintuitive.

\begin{figure}[ttt]
\begin{center}
\includegraphics[height=8cm,width=6cm,angle=270]{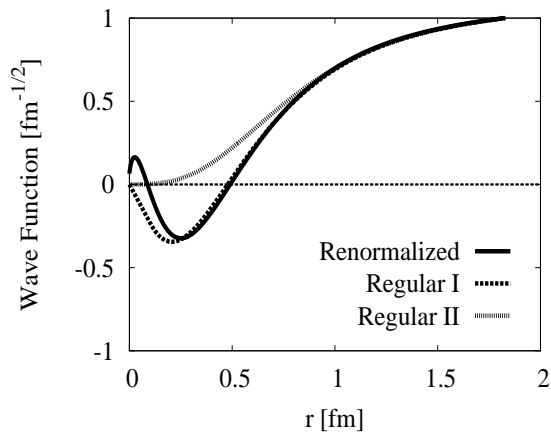}
\end{center}
\caption{Zero energy wave function for the singlet $np$ $^1S_0$ channel as a
function of distance (in fm) and for the different scenarios with large and
small $\omega-$couplings. This wave function goes asymptotically to $u_0(r)
\to 1 - r /\alpha_0$ with $\alpha_0 = -23.74 {\rm fm}$ the scattering length in
this channel. The zero at about $r=0.5 {\rm fm}$ signals the existence of a
spurious bound state.}
\label{fig:u0-r-CSB}
\end{figure}

However it is worth mentioning that the different scenarios correspond
to selecting a potential possessing spurious bound states or not. The
spurious bound state problem has been discussed in
Ref.~\cite{Cordon:2009pj} at length; the number of inner zeroes of the
zero energy wave function provides the number of bound states.  For
illustration we represent the zero energy wave in
Fig~\ref{fig:u0-r-CSB}.  In the regular case, the OBE potential with a
big $g_{\omega NN}^*$ is free of bound states.  However if a small
$g_{\omega NN}^*$ is chosen, then one has to deal with a bound state
which does not exist and it is hence spurious. 

\subsection{Irregular solutions}

The previous results are in conflict with the intuitive expectation of
insensitivity of low energy physical observables with respect to the
specific details of the potential in the short distance
region. Otherwise, where should one stop ?. This is the basis of the
renormalization viewpoint.
The way to proceed is to impose renormalization conditions which
eliminate the short range sensitivity at the expense of treating low
energy parameters as independent variables from the potential.
An example of a renormalization condition (RC) is to fix the scattering length,
with the consequence of avoiding the fine tuning problem summarized by
Eq.~(\ref{eq:delta-alpha0}).
In other words, we trade the explicit dependence of the results on
the short range parameters of the potential for low energy observables.
The values of the later are usually well-known by other means.

In principle there are several ways in which one can impose renormalization
conditions, one popular example being counterterms.
They correspond to the coupling constants of a short distance contact 
potential which is expanded in terms of $\delta$ functions and its derivatives
\begin{eqnarray}
V_C(\vec{x}) = C_0\,\delta(\vec{x}) + C_2\,\{\nabla^2, \delta(\vec{x})\} +
\dots \, ,
\end{eqnarray}
where the dots represent terms involving higher derivatives
of the $\delta$ function.
This potential is added to the usual finite range potential $V$,
and then the corresponding Schr\"odinger equation for $V_C + V$ is solved.
The resulting potential is strongly singular and needs to be regularized
by introducing a cut-off $r_c$, a length scale which is used to smear
the $\delta$ functions.
The different coupling constants $C_0(r_c)$, $C_2(r_c)$, etc, are set
to reproduce the desired low-energy observables.
A nice presentation of the previous method is given in
Ref.~\cite{Lepage:1997cs}.
The disadvantage is that the procedure of using a potential to renormalize
quickly runs into problems when one tries to decrease the size of the cut-off.
For example, it may be impossible to reproduce certain physical
observables, specifically the effective range, if the short distance
cut-off is too small~\cite{Phillips:1996ae}, unless one accepts
complex values for the counterterms $C_0(r_c)$ and $C_2(r_c)$ which
violate either causality or off-shell unitarity (see
Ref.~\cite{Entem:2007jg} for a detailed discussion). Related
positivity conditions are discussed in Ref.~\cite{Cordon:2009pj}.

Here we use a more indirect method to renormalize which is able to avoid
some of the previously mentioned complications.
The idea is to substitute the regularity condition of the Schr\"odinger
equation, $u_k(0) = 0$, by an arbitrary boundary condition at the origin
\begin{eqnarray}
L_k(0) = \frac{u_k'(0)}{u_k(0)} \, ,
\end{eqnarray}
where we have used the log-derivative of the wave function, instead of an
independent condition for $u_k(0)$ and $u_k'(0)$, as that will only affect
the precise normalization of the wave function, 
which can be later determined by other means.
The regularity condition $u_k(0) = 0$ corresponds to taking the limit
$L_k(0) \to \infty$ (as $u_k'(0)$ is a constant),
but by changing the precise value and energy dependence $L_k(0)$,
the values of low energy observables can be fixed.

The previous procedure would in principle involve a fitting procedure,
which can be avoided by taking into account the expansion in powers of
$k$ of the wave function, Eq.~(\ref{eq:uk_exp}).
For example, if we want to fix the scattering length, we will solve
the corresponding equation for the zero-energy wave function $u_0(r)$,
with the asymptotic ($r \to \infty$) boundary condition of reproducing
the scattering length
\begin{eqnarray}
-u_0''(r) + M_N\,V(r) u_0(r) &=& 0 \, , \\
u_0(r) &\to& 1 - \frac{r}{\alpha_0}, 
\end{eqnarray}
but, instead of solving the previous equation from $r=0$ to $r \to \infty$,
we solve it downwards from infinity to the origin.
Then, we assume that $u_2(r)$, $u_4(r)$, etc, are subjected to regular
boundary conditions at the origin, $u_2(0) = u_4(0) = 0$ and
$u_2'(0) = u_4'(0) = 0$, which means to take 
\begin{eqnarray}
L_k(0) = \frac{u_0'(0)}{u_0(0)} \, .
\end{eqnarray}
By so doing we achieve some insensitivity at short distances, as we
will show later~\footnote{In dispersion theory these renormalization
  condition resembles the customary subtractions. In our case the form
  of the subtraction is a bit more involved as discussed below.}.
Fixing more scattering parameters is straightforward: one solves downwards
the corresponding equations for $u_0(r), u_2(r), \dots, u_{2n}(r)$
with the asymptotic conditions of reproducing
$\alpha_0, r_0, \dots, v_{n}$, 
and assumes trivial boundary conditions for $u_{2n+2}(r)$, $u_{2n+4}(r)$, etc,
resulting the following logarithmic boundary condition
\begin{eqnarray}
L_k(0) = 
\frac{u_0'(0) + k^2 u_2'(0) + \dots + k^{2n} u_{2n}'(0)}
{u_0(0) + k^2 u_2(0) + \dots + k^{2n} u_{2n}(0)}\, .
\end{eqnarray}
In practical computations it is convenient to introduce a short distance
cut-off, $r_c$, and then take the limit $r_c \to 0$.

A further simplification can be made if we take into account that
the OBE potential of Eq.~(\ref{eq:potential-OBE}) is local and
energy independent.
This means in particular that different energy states are orthogonal,
\begin{eqnarray}
\label{eq:orthogonality-int}
\int_0^\infty u_k (r) u_p (r)  dr = 0\, , 
\end{eqnarray} 
for $k \neq p$, which requires an energy independent boundary condition
at the origin, as a consequence of the next equality
\begin{eqnarray}
\int_0^\infty u_k (r) u_p (r)  dr = u_k' u_p - u_p u_k' \Big|_0 \, , 
\end{eqnarray}
which means that the orthogonality condition Eq.~(\ref{eq:orthogonality-int})
can be re-expressed as
\begin{eqnarray}
\frac{u_k'(0)}{u_k(0)} = \frac{u_p'(0)}{u_p(0)} \, ,
\end{eqnarray}
or, equivalently, $L_k(0) = L_p(0)$, implying an energy independent
boundary condition.

Here we will only consider the case in which orthogonality is preserved.
The restriction is that orthogonality implies that we can only fix
one scattering parameter, namely the scattering length.
Therefore we will integrate downwards the zero energy state,
$u_0 (r) \to 1- {r}/{\alpha_0}$, up to the cut-off $r_c$.
At this point, we can make use of the superposition principle in order
to take real advantage of the boundary condition method.
For the zero-energy solution, the superposition principle can be used
to write the wave function as a linear combination of two independent
zero-energy solutions
\begin{eqnarray}
\label{eq:u0-superposition}
u_0(r) = u_1(r) - \frac{u_r(r)}{\alpha_0} \, ,
\end{eqnarray}
where $u_1(r)$ and $u_r(r)$ are solutions of the zero-energy Schr\"odinger
equation, Eq.~(\ref{eq:u0}), with the asymptotic boundary conditions
\begin{eqnarray}
u_1(r) &\to& 1 \, , \\
u_r(r) &\to& r \, ,
\end{eqnarray}
at large distances.
The previous expression for $u_0$ as a linear combination,
Eq.~(\ref{eq:u0-superposition}),
can be introduced in the integral representation of the effective range $r_0$,
Eq.~(\ref{eq:r0}),
yielding the following correlation between $r_0$ and $\alpha_0$
\begin{eqnarray} 
r_0  &=&  A_0 + \frac{B_0}{\alpha_0}+ \frac{C_0}{\alpha_0^2}   \, , 
\label{eq:r0_univ} 
\end{eqnarray} 
where $A_0$, $B_0$ and $C_0$ are given by
\begin{eqnarray}
A_0 &=& 2 \int_0^\infty dr ( 1 - u_{1}^2 ) \, , \\    
B_0 &=& -4 \int_0^\infty dr ( r - u_{1} u_{r} ) \, , \\    
C_0 &=& 2 \int_0^\infty dr ( r^2 - u_{r}^2 ) \, .
\end{eqnarray} 
The interesting feature is that the dependence of the effective range
with respect to short-range parameters of the potential is greatly
diminished.

\begin{figure}[tbc]
\begin{center}
\includegraphics[height=7.5cm,width=6.5cm,angle=270]
{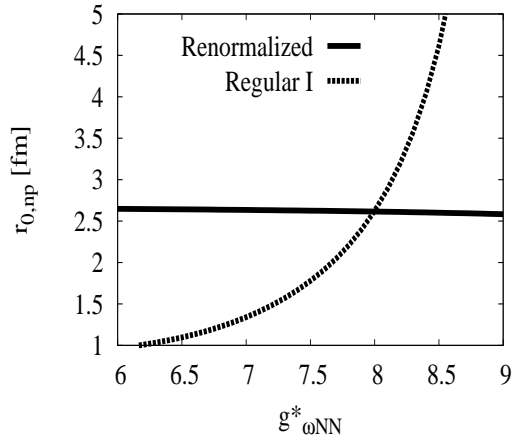}
\end{center}
\caption{Dependence of the effective range with respect to $g_{\omega NN}^*$
in the regular case with a small coupling constant and in the renormalized one.}
\label{fig:r0-renormalized}
\end{figure}

The short distance sensitivity can be vividly seen in
Fig.~\ref{fig:r0-renormalized}, where the regular (parabola like
curve) as well as the renormalized (flat curve) effective range for
the OBE potential are shown as a function of $g_{\omega NN}^*$. For
simplicity only the solution with the small $g_{\omega NN}^*$ (Regular
solution I) is represented.

The finite-energy solutions and the phase shifts can be obtained 
from the orthogonality condition, which implies
\begin{eqnarray}
\label{eq:ortho-Lk}
\lim_{r_c \to 0} \frac{u_k' (r_c)}{u_k(r_c)} =
\lim_{r_c \to 0} \frac{u_0' (r_c)}{u_0(r_c)} \, ,
\end{eqnarray}
providing the initial boundary conditions for the finite-energy
Schr\"odinger equation, Eq.~(\ref{eq:schroe-s}).
We normalize the scattering wave function as follows
\begin{eqnarray}
u_k (r) \to \frac{\sin(kr + \delta_0)}{\sin \delta_0} \, .
\end{eqnarray} 
Again, if we use the superposition principle, $u_k$ can be written as
\begin{eqnarray}
u_k (r) = k \cot \delta_0 \, u_{k,s} (r) + u_{k,c} (r) \, , 
\end{eqnarray} 
with $u_{k,c}$ and $u_{k,s}$ solutions of Eq.~(\ref{eq:schroe-s})
asymptotically normalized as
\begin{eqnarray}
u_{k,c} &\to& \cos (k r) \, , \\
u_{k,s} &\to& \frac{\sin (k r)}{k} \, .
\end{eqnarray}
These two wave functions have the property of going to $u_{k,s} \to u_r$
and $u_{k,c} \to u_1$ for $k \to 0$.
Then, the orthogonality constraint Eq.~(\ref{eq:ortho-Lk}) reads
\begin{eqnarray}
\frac{k \cot \delta_0 u_{k,s}' (r_c) + u'_{k,c} (r_c)}
{k \cot \delta_0 u_{k,s} (r_c) + u_{k,c} (r_c)} = 
\frac{\alpha_0 \, u_{1}' (r_c) - u_{r}' (r_c)}
{\alpha_0 \, u_{1} (r_c) - u_{r} (r_c)} \, ,
\end{eqnarray} 
where we have dropped the $\lim_{r_c \to 0}$ for shortening the notation.
Note that the dependence of the phase-shift on the scattering length
is explicit: $k \cot \delta_0 $ is a bilinear mapping of $\alpha_0$, 
\begin{eqnarray}
k \cot \delta_0 =
\frac{ \alpha_0 {\cal A} ( k) + {\cal B} (k)}
{ \alpha_0 {\cal C} ( k) + {\cal D} (k)} \, ,
\label{eq:phase_singlet}
\end{eqnarray} 
where the functions ${\cal A}$, ${\cal B}$, ${\cal C}$ and ${\cal D}$
are even functions of $k$ which depend solely on the potential
and are given by the following formulas 
\begin{eqnarray}
{\cal A}(k) &=& \lim_{r_c \to 0} 
\left( u_{1} (r_c)  u'_{k,c} (r_c) - 
u_{1}' (r_c)  u_{k,c} (r_c) \right)  \, , \nonumber \\  
{\cal B}(k) &=& \lim_{r_c \to 0} \left( u_{k,c} (r_c)  u'_{r} (r_c) - 
u_{r} (r_c)  u_{k,c}' (r_c) \right)  \, , \nonumber \\  
{\cal C}(k) &=& \lim_{r_c \to 0} \left( u_{1}' (r_c)  u_{k,s} (r_c) - 
u_{1} (r_c)  u'_{k,s} (r_c) \right)  \, , \nonumber \\  
{\cal D}(k) &=& \lim_{r_c \to 0} \left( u_{r} (r_c)  u'_{k,s} (r_c) - 
u_{r}' (r_c)  u_{k,s} (r_c) \right)  \, . \nonumber \\ 
\end{eqnarray}
The previous expressions fix the arbitrary normalization of
Eq.~(\ref{eq:phase_singlet}).
The obvious conditions ${\cal A}(0)={\cal D}(0)=0$  and  
${\cal B}(0)={\cal C}(0)=1$  are satisfied.
Expanding the expression  for small $k$ one gets that $v_k$
is a polynomial in $1/\alpha_0 $ of degree $k+1$. 

Finally, we can use the previous procedure to fit the $np$ averaged
data of Ref.~\cite{Stoks:1994wp} in the $^1S_0$ channel (once we have
fixed the scattering length to its experimental value), yielding the
values in table \ref{tab:table_fits_CSB}.  We can see the large
uncertainty on the value of $g_{\omega NN}^*$, which shows that there
is a greater insensitivity to shorter distances after renormalizing.
This agrees with the previous remarks on the sensitivity of the
effective range on $g_{\omega NN}^*$, illustrated in
Fig.~\ref{fig:r0-renormalized}. Let us note further that as discussed
in Ref.~\cite{Cordon:2009pj} the renormalization scenario also has
also a spurious bound state as in the small $g_{\omega NN}^*$ regular
solution case (see Fig~\ref{fig:u0-r-CSB}). The current discussion
would be modified by the inclusion of form factors which incorporate
the finite nucleon size. However, because of the short distance
insensitivity form factors turn out to play a marginal
role~\cite{Cordon:2009pj} {\it after renormalization} .

\subsection{Review of the Renormalization Process}

The renormalization procedure proposed in this section
can be summarized as follows
\begin{itemize}  
\item 
For a given scattering length $\alpha_0$, integrate in the zero energy wave
function $u_0(r)$ with Eq.~(\ref{eq:u0}) down to the cut-off radius $r_c$.
This is the renormalization condition.

\item
Implement self-adjointness at the cut-off radius through the boundary condition
\begin{eqnarray} 
u_k'(r_c)  u_0 (r_c) - u_0'(r_c) u_k(r_c) =0 \, , 
\label{eq:boundary}
\end{eqnarray} 

\item
Integrate out the finite energy wave function $u_k(r)$
with Eq.~(\ref{eq:schroe-s})
and  determine the phase shift $\delta_0(p)$ from Eq.~(\ref{eq:d0}). 

\item
Remove the cut-off (take the limit $r_c \to 0$)
to assure model (regulator) independence.
\end{itemize}
This allows to compute $\delta_0$ (and hence $r_0$, $v_2$ ) from
(i) the potential $V(r)$ and (ii) the scattering length $\alpha_0$ as 
{\it independent} information. 
Note that this is equivalent to consider, in addition to the regular solution,
the {\it irregular} one. In momentum space this can be shown to
be equivalent to introduce one counterterm in the cut-off Lippmann-Schwinger
equation (see Ref.~\cite{Entem:2007jg} for a detailed discussion).
Both Eq.~(\ref{eq:r0_univ}) and Eq.~(\ref{eq:phase_singlet}) highlights
this de-correlation between the potential and the scattering length.
Contrary to common wisdom, but according to our intuitive
expectations, no strong short range repulsion is essential.
The moral is that building $\alpha_0$ {\it from} the potential is equivalent
to absolute knowledge at short distances,
as in the $^1S_0$ channel a strong fine tuning is at work. 
This example illustrates our point that the renormalization viewpoint
tells us to what extent short distance physics may be less well determined
than the traditional approach assumes.
This opens up a new perspective (see \cite{Cordon:2009pj}) to the
phenomenology of OBE potentials in cases where the strong
$\omega$-repulsion has proven to be crucial at low energies.

\section{Renormalization with Coulomb interactions}
\label{sec:Coulomb}

In this section we generalize the previously discussed renormalization approach
to the case of proton-proton scattering, where the infinite range of
the Coulomb interaction will pose some problems. 
The corresponding $s-$wave reduced Schr\"odinger equation is
\begin{eqnarray}
\label{eq:schroe_uk_C}
-{u_k^C}'' + M_p\,\left( V_{pp} (r) + \frac{\alpha}{r} \right)
\,u_k^C = k^2\,u_k^C \, ,
\end{eqnarray}
where $k$ is the center-of-mass momentum, $M_p$ the proton mass, $V_{pp}(r)$
the strong proton-proton potential, and $\alpha \simeq {1}/{137}$ is
the fine structure constant.
Actually, the current discussion is tightly linked to the corresponding one
for the two potential formula presented by two of us
recently~\cite{PavonValderrama:2009nn}.

\subsection{Coulomb scattering at zero energy}

The longest range piece of the proton-proton interaction is the
Coulomb repulsion between the protons.
Ignoring any strong effects, zero energy proton-proton scattering in $s-$waves
can be described by the reduced Schr\"odinger equation
\begin{eqnarray}
\label{eq:schroe_v0_C}
-{v_0^C}'' + \frac{2}{a_B r}\,v_0^C(r) = 0 \, ,
\end{eqnarray}
where $a_B$ is the proton Bohr radius, which is defined as
$a_B = 2 / M_p \alpha = 56.62\,{\rm fm}$.
The previous equation has the following two linearly independent solutions
\begin{eqnarray}
v^C_{0,\rm reg}(r) &=& \frac{a_B}{2}\,\sqrt{x}\,I_1(2\,\sqrt{x}) \, , \\
v^C_{0,\rm irr}(r) &=& 2\,\sqrt{x}\,K_1(2\,\sqrt{x}) \, ,
\end{eqnarray}
where $x = 2\,r / a_B$ and $K_1(x)$ and $I_1(x)$ are modified Bessel
functions of the first and second kind respectively
(see for example~\cite{abramowitz+stegun}).
At short distances these solutions behave as
\begin{eqnarray}
\label{eq:v_C_short}
v^C_{0,\rm reg}(r) &\to& r + \frac{r^2}{a_B} + \frac{r^3}{3\,a_B^2}
+ {\mathcal O}(r^4)\, , \\
v^C_{0,\rm irr}(r) &\to& 1 + \frac{2 r}{a_B}\,
\left[ \log{\frac{2 r}{a_B}} + 2 \gamma_E - 1 \right] \nonumber\\ 
&+&
{\left(\frac{2 r}{a_B}\right)}^2\,
\left[ \frac{1}{2}\,\log{\frac{2 r}{a_B}} + \gamma_E - \frac{5}{4}\right]
+ {\mathcal O}(r^3)\, ,
\nonumber \\
\end{eqnarray}
where $\gamma_E = 0.57722$ is the Euler-Mascheroni constant. 
The previous means in particular that $v^C_{0,\rm reg}$ is
the short distance regular solution and $v^C_{0,\rm irr}$ 
the short distance irregular solution.
In principle, in the absence of any strong potential, $v^C_{0,\rm reg}$
would be the zero energy solution for the repulsive Coulomb potential.
The presence of the strong interaction between the protons means that
the zero-energy asymptotic solution for $r \to \infty$ will be
in general a linear combination of $v^C_{0,\rm reg}$ and $v^C_{0,\rm irr}$.

For the proton-proton system the Coulomb scattering length is related with
the asymptotic behaviour of the zero energy wave function
at large enough distances
\begin{eqnarray}
\label{eq:v_0C}
v^C_{0}(r) = v^C_{0,irr}(r) - \frac{v^C_{0,reg}(r)}{\alpha_{0,C}} \, ,
\end{eqnarray}
where $v^C_{0,\rm reg}$ and $v^C_{0,\rm irr}$ are the previously defined
regular and irregular zero energy wave functions, and $\alpha_{0,C}$ is
the $s-$wave Coulomb scattering length.
If the Coulomb interaction is switched off by taking $a_B \to \infty$,
the previous wave function reduces to the corresponding one for finite
range forces, $v(r) = 1 - r / \alpha_0$.

\subsection{Effective range}

The Coulomb effective range is given by the following formula
\begin{eqnarray}
\label{eq:r0C-int}
r_{0,C} = 2\,\int_0^\infty \, dr \left[ v^C_0 (r)^2 - u^C_0 (r)^ 2 \right] \, , 
\end{eqnarray}
where $v^C_0 (r)$ is the Coulomb zero energy solution given by
Eq.~(\ref{eq:v_0C}), and $u^C_0 (r)$ is the {\it full}
zero energy solution to the Schr\"odinger equation
\begin{eqnarray}
\label{eq:schroe_u0_C}
-{u_0^C}'' + 
\left( M_p\,V_{pp}(r) + \frac{2}{a_B r} \right) \,u_0^C(r) = 0 \, ,
\end{eqnarray}
subjected to the asymptotic boundary condition
\begin{eqnarray}
\label{eq:u_0C}
u^C_0(r) \to v^C_0(r) \, ,
\end{eqnarray}
for $r \to \infty$.
By making use of the superposition principle, the solution $u^C_0$
can be decomposed as
\begin{eqnarray}
\label{eq:u_0C-s} 
u^C_0(r) = u^C_{0,\rm irr}(r) - \frac{u^C_{0,\rm reg}(r)}{\alpha_{0,C}} \, ,
\end{eqnarray}
where $u^C_{0,\rm irr}$ and $u^C_{0,\rm reg}$ are solutions of the zero energy
Schr\"odinger equation,  Eq.~(\ref{eq:schroe_u0_C}),
behaving asymptotically ($r \to \infty$) as
\begin{eqnarray}
u^C_{0,\rm reg}(r) &\to& v^C_{0,\rm reg}(r) \, , \label{eq:u_0C_reg_def} \\
u^C_{0,\rm irr}(r) &\to& v^C_{0,\rm irr}(r) \, . \label{eq:u_0C_irr_def}
\end{eqnarray}
The subscripts ${}_{\rm reg}$ and ${}_{\rm irr}$ do not refer to the regularity
of the solutions at the origin, but with the long range behaviour of the full
solutions.

By plugging the decomposition of the full and purely Coulomb wave functions,
Eqs.~(\ref{eq:u_0C}) and (\ref{eq:v_0C}),
into the integral representation of the Coulomb effective range, 
Eq.~(\ref{eq:r0C-int}),
we obtain the following correlation between the Coulomb scattering length
and effective range
\begin{eqnarray}
r_{0,C} &=& A_0^C + \frac{B_0^C}{\alpha_{0,C}} + \frac{C_0^C}{\alpha_{0,C}^2} \, , 
\end{eqnarray}
which is a direct generalization of Eq.~(\ref{eq:r0_univ})
for the non-Coulomb case.
The Coulomb correlation functions $A_0^C$, $B_0^C$ and $C_0^C$ are given by
the integral expressions below
\begin{eqnarray}
A_0^C &=&  2\,\int_0^\infty dr ( {v^C_{0,\rm irr}(r)}^2 - 
{u^C_{0,\rm irr}(r)}^2 ) \, , \\    
B_0^C &=& -4\,\int_0^\infty dr ( v^C_{0,\rm irr}(r)\,v^C_{0,\rm reg}(r)
- u^C_{0,\rm irr}(r)\,u^C_{0,\rm reg}(r)) \, , \nonumber \\
&& \\    
C_0^C &=&  2\,\int_0^\infty dr ( {v^C_{0,\rm reg}(r)}^2 - 
{u^C_{0,\rm reg}(r)}^2 ) \, .
\end{eqnarray} 

\subsection{Coulomb scattering at finite energy and Coulomb effective range expansion}

The definition of the phase shifts in the presence of the Coulomb
potential is related with the behaviour of the wave function at long
distances, which is given by
\begin{eqnarray}\label{eq:ukC_cotd}
u^{C}_k(r) \to \cot{\delta^{C}_0(k)} F_0(\eta, \rho) + G_0(\eta, \rho) \,  
\end{eqnarray}
where $\delta^{C}_0(k)$ is the Coulomb-modified proton-proton phase shift,
$k$ the center-of-mass momentum and $F_0(\eta, \rho)$ and $G_0(\eta, \rho)$,
with $\eta  = 1 / k a_B$ and $\rho = k r$, are the $s-$wave Coulomb wave
functions~\cite{abramowitz+stegun}.
The $u^{C}_k$ wave function is the solution to the reduced Schr\"odinger
equation Eq.~(\ref{eq:schroe_uk_C}).
The $F_0(\eta, \rho)$ and $G_0(\eta, \rho)$ wave functions behave
asymptotically ($r \to \infty$) as
\begin{eqnarray}
F_0 &\to& \sin{\left( k r - \eta \log(2 k r) + \sigma_0 \right)}  \, , \\
G_0 &\to& \cos{\left( k r - \eta \log(2 k r) + \sigma_0 \right)}  \, ,
\end{eqnarray}
with $\sigma_0$ the Coulomb phase shift, which is defined as
\begin{eqnarray}
e^{2 i \sigma_0} = \frac{\Gamma(1 + i \eta)}{\Gamma(1 - i \eta)} \, .
\end{eqnarray}

The phase shift in presence of the infinite-ranged Coulomb force
does not obey the usual effective range expansion, 
valid for short-ranged potentials, but a Coulomb-modified effective range
expansion, given by
\begin{eqnarray}
k\,\cot{\delta^{C}_0}\,C^2(\eta) + \frac{2}{a_B}\,h(\eta) 
&=& - \frac{1}{\alpha_{0,C}} + \frac{1}{2}\,r_{0,C}\,k^2 \nonumber \\
&+& \sum_{n=2}^{\infty}\,v_{n,C} k^{2n} \, ,
\label{eq:ERE-Coulomb}
\end{eqnarray}
with $C(\eta)$ and $h(\eta)$ defined as
\begin{eqnarray}
C^2(\eta) &=& \frac{2 \pi \eta}{e^{2 \pi \eta} - 1} \, , \\
h(\eta) &=& \eta^2\,\sum_{n=1}^{\infty}\,\frac{1}{n (n^2 + \eta^2)} - 
\log{\eta} - \gamma_E \, .
\end{eqnarray}

For obtaining the Coulomb extension of Eq.~(\ref{eq:phase_singlet}) we use
again the superposition principle to write $u^C_k$ in the following way
\begin{eqnarray}
C(\eta)\,u^C_k(r) &=&
\left( k\,\cot{\delta^{C}_0}\,C^2(\eta) + \frac{2}{a_B}\,h(\eta) 
\right)\,u^C_{k,\rm reg}(r) \nonumber \\ &-& u^C_{k,\rm irr}(r) \, ,
\end{eqnarray}
where $u^C_{k,\rm reg}$ and $u^C_{k,\rm irr}$ are solutions of
Eq.~(\ref{eq:schroe_uk_C}), which obey the asymptotic boundary conditions
\begin{eqnarray}
u^C_{k,\rm reg}(r) &\to& \frac{F_0(\eta, \rho)}{k C(\eta)} \, , \\
u^C_{k,\rm irr}(r) &\to& - C(\eta)\,G_0(\eta, \rho) +
\frac{2 \eta h(\eta)}{C(\eta)}\,F_0(\eta, \rho)\, , \nonumber \\
\end{eqnarray}
for $r \to \infty$.
These two solutions have been normalized in such a way that for $k \to 0$
they coincide with the previously defined zero-energy wave functions
$u^C_{0,\rm reg}$ and $u^C_{0,\rm irr}$,
see Eqs.~(\ref{eq:u_0C_reg_def}) and (\ref{eq:u_0C_irr_def}).

Once these definitions have been made, it is straightforward
to obtain the correlation
\begin{eqnarray}
k\,\cot{\delta^{C}_0}\,C^2(\eta) + \frac{2}{a_B}\,h(\eta) =
\frac{ \alpha_{0,C} {\cal A}^C ( k) + {\cal B}^C (k)}
{ \alpha_{0,C} {\cal C}^C ( k) + {\cal D}^C (k)} \, ,
\nonumber \\
\label{eq:phase_singlet_coulomb}
\end{eqnarray}
where ${\cal A}^C (k)$, ${\cal B}^C (k)$, ${\cal C}^C (k)$ and ${\cal D}^C (k)$
are defined as
\begin{eqnarray}
\label{eq:A}
{\cal A}^C (k) = \lim_{r_c \to 0}
&\Big[& u^C_{0,\rm irr} (r_c)  {u^C_{k,\rm irr}}' (r_c) \nonumber \\ 
&-& {u^C_{0,\rm irr}}' (r_c)  u^C_{k,\rm irr} (r_c) \Big] \, , \\
\label{eq:B} 
{\cal B}^C (k) = \lim_{r_c \to 0} 
&\Big[& u^C_{0,\rm reg} (r_c)  {u^C_{k,\rm irr}}' (r_c) \nonumber \\
&-& u^C_{k,\rm irr} (r_c)  {u^C_{0,\rm reg}}' (r_c) \Big] \, , \\ 
\label{eq:C}  
{\cal C}^C (k) = \lim_{r_c \to 0} &\Big[& 
u^C_{0,\rm irr} (r_c) {u^C_{k,\rm reg}}' (r_c) \nonumber \\
&-& {u^C_{0,\rm irr}}' (r_c)  u^C_{k,\rm reg} (r_c) \Big] \, , \\ 
\label{eq:D}  
{\cal D}^C (k) = \lim_{r_c \to 0} &\Big[& 
u^C_{0,\rm reg} (r_c) {u^C_{k,\rm reg}}' (r_c)  \nonumber \\
&-& {u^C_{0,\rm reg}}' (r_c)  u^C_{k,\rm reg} (r_c) \Big] \, . 
\end{eqnarray}

\section{Charge symmetry breaking}
\label{sec:CSB}

In the previous sections we have shown how the renormalization of the
$^1S_0$ two nucleon system can be carried out.
This procedure allows to determine the $^1S_0$ phase shifts for
$np$, $nn$, $pp$ and $pp(c)$ from their corresponding scattering lengths
$\alpha_{np}$, $\alpha_{nn}$, $\alpha_{pp}$ and $\alpha_{pp}^C$ respectively.
The previous computation can be compared with the experimental values for
these quantities in order to test the renormalization procedure.

Admitted values of the scattering lengths are~\cite{Machleidt:2001rw}, 
\begin{eqnarray}
\alpha_{0,pp}^C = -7.8149(29)\,{\rm fm}\, , && \alpha_{0,pp} = -17.3(4)\,{\rm fm}
\, , \nonumber \\  
\alpha_{0,nn}  = -18.8(3)\,{\rm fm} \,\, \,  , && \alpha_{0,np} = -23.77(9)\,{\rm fm}
\, , \nonumber \\  
\end{eqnarray}
giving $\Delta \alpha_{CIB} = 5.7\,{\rm fm}$ and
$\Delta \alpha_{CSB} = 1.5\,{\rm fm}$.
For the effective range we have~\cite{Machleidt:2001rw}, 
\begin{eqnarray}
r_{pp}^C = 2.769(14)\,{\rm fm} \, , &&  r_{pp} = 2.85(4)\,{\rm fm}
\, , \nonumber \\  
r_{nn} = 2.75(11)\,{\rm fm} \, \, \, , && r_{np} = 2.75(5)\,{\rm fm}
\, , \nonumber \\  
\end{eqnarray}
with  $\Delta r_{0,CIB} = 0.05\,{\rm fm}$ and
$\Delta r_{0,CSB} = 0.1\,{\rm fm}$.
As can be seen, the CIB/CSB breaking is much larger for the scattering length
than the effective range.
Part of that is explained by the unnaturally large value
of the $NN$ scattering length.
\begin{figure*}[ttt]
\includegraphics[height=7cm,width=5.5cm,angle=270]{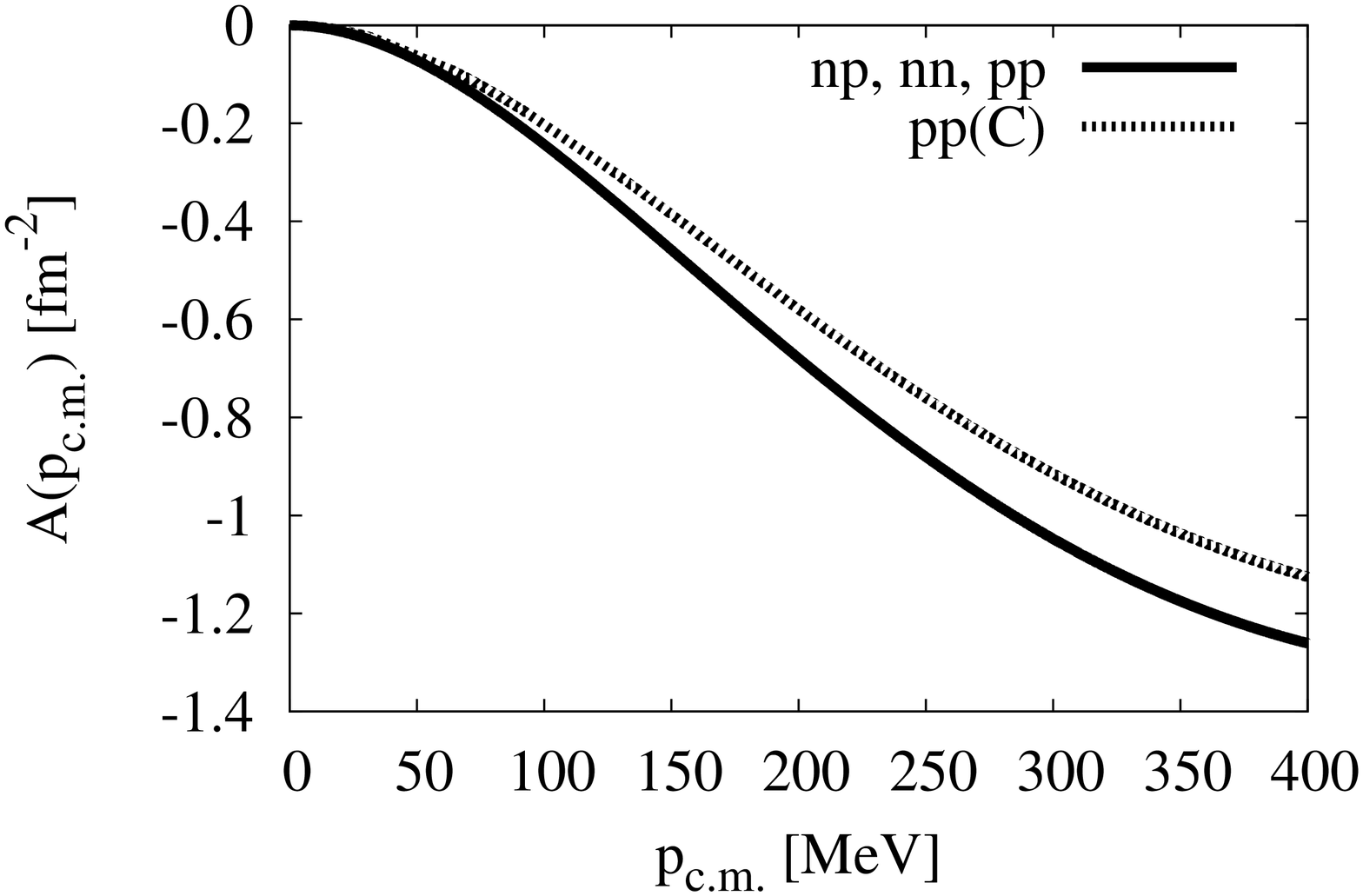}
\includegraphics[height=7cm,width=5.5cm,angle=270]{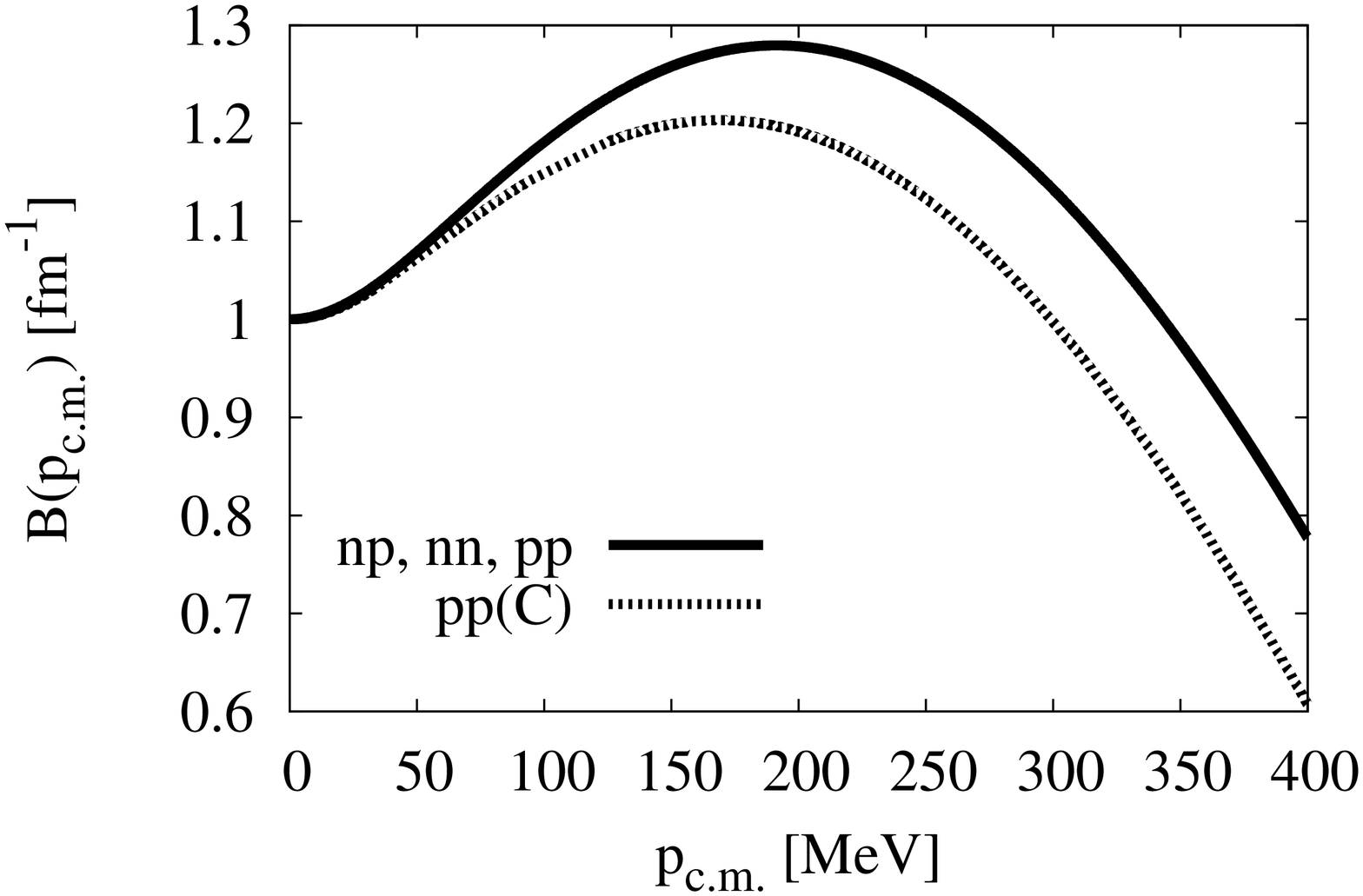} \\
\includegraphics[height=7cm,width=5.5cm,angle=270]{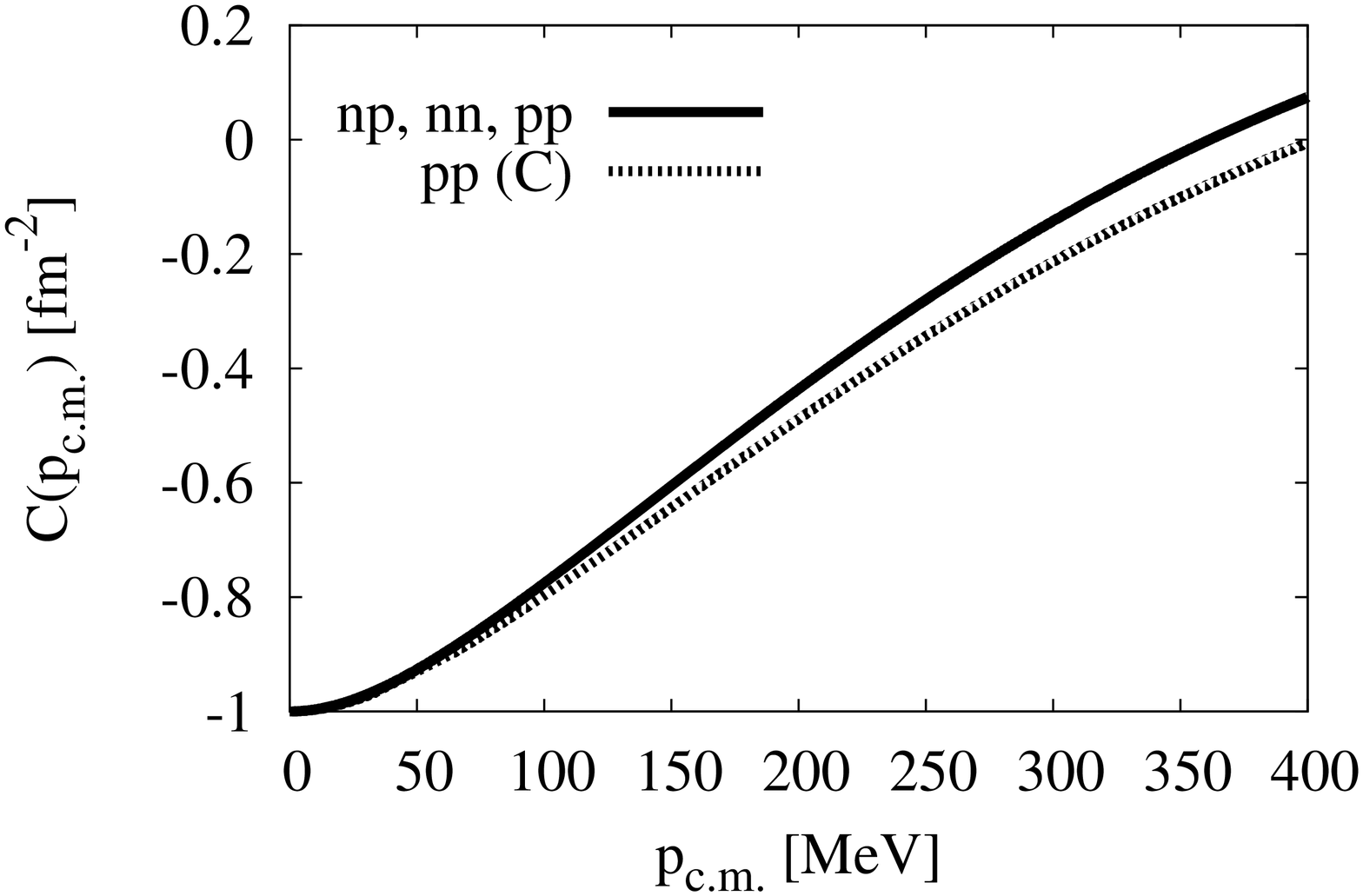}
\includegraphics[height=7cm,width=5.5cm,angle=270]{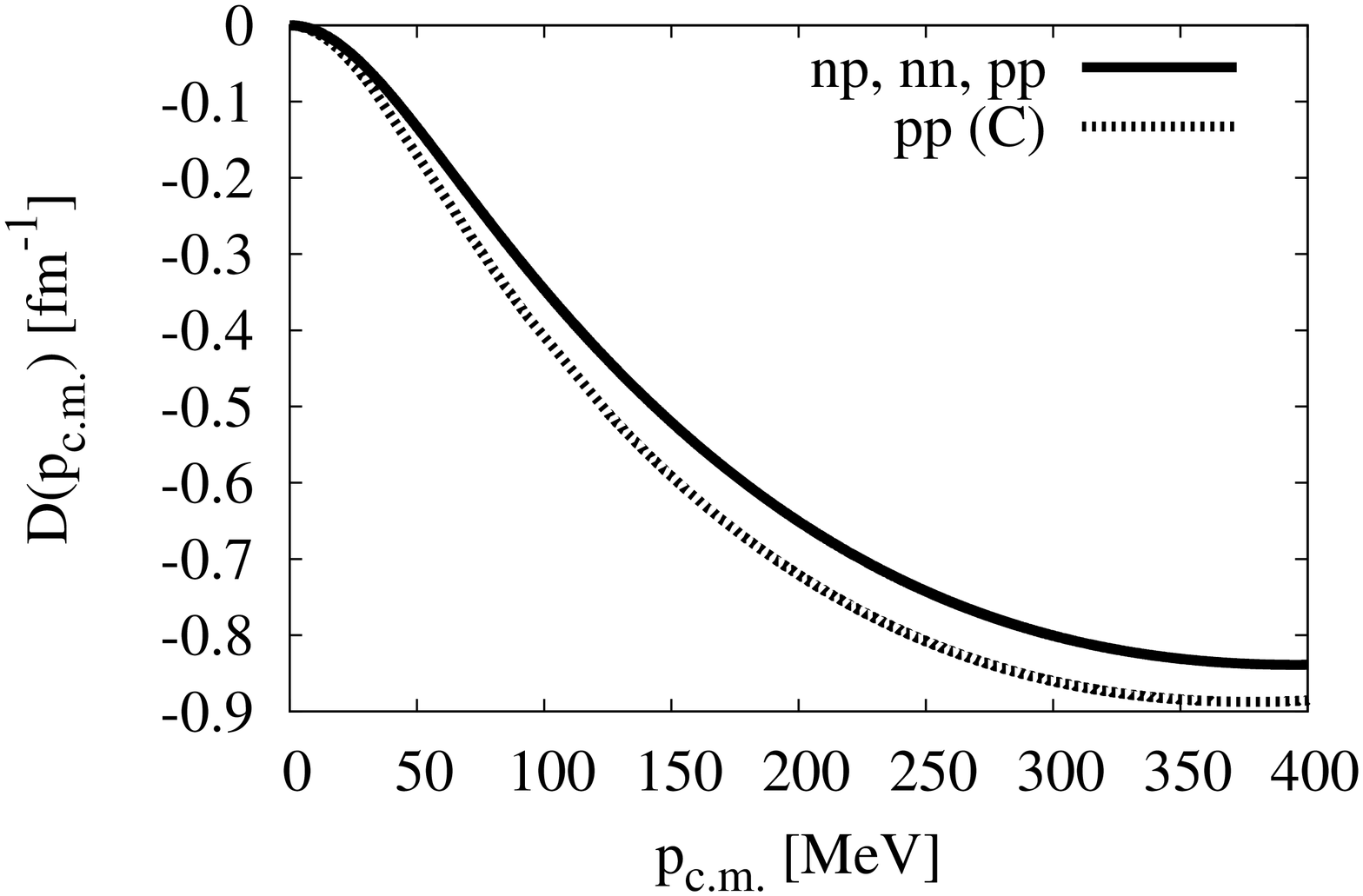}
\caption{The universal functions ${\cal A}$, ${\cal B}$ ${\cal C}$ and
  ${\cal D}$ defined by Eqs.~(\ref{eq:A}-\ref{eq:D})
  in appropriate length
  units as a function of the CM momentum $p$ (in MeV) for the four
  $^1S_0$ channels $np$,$pp$,$nn$,$pp(c)$. These functions depend on the
  potentials $V_{np} (r) $, $V_{nn} (r) $, $V_{pp} (r) $, $V_{pp}^C
  (r) $ only but are independent of the scattering lengths.}
\label{fig:ABCD}
\end{figure*}

To take into account the various physical effects which generate
charge symmetry breaking,
we consider the neutron-proton mass difference and the OPE potentials
\begin{eqnarray}
U_{pp}^{1\pi} (r) &=& - M_p \,f_{\pi}^2\,
{\left(\frac{m_{\pi^0}}{m_{\pi^+}}\right)}^2\,
\frac{e^{-m_{\pi_0} r}}{r } \, , \nonumber \\ 
U_{nn}^{1\pi} (r) &=& - M_n \,f_{\pi}^2\,
{\left(\frac{m_{\pi^0}}{m_{\pi^+}}\right)}^2\,
\frac{e^{-m_{\pi_0} r}}{r } \, , \nonumber \\ 
U_{np}^{1\pi} (r) &=& -M_{np}\,f_{\pi}^2\,
\left[ 2\,  \frac{e^{-m_{\pi_+} r}}{r } -
{\left(\frac{m_{\pi^0}}{m_{\pi^+}}\right)}^2\,
\frac{e^{-m_{\pi_0} r}}{r } \right]
\, , \nonumber \\ 
\end{eqnarray} 
with $m_{\pi_0}= 134.97\,{\rm MeV}$ and $m_{\pi_+}= 139.57\,{\rm MeV}$.
$M_{np}$ is twice the $np$ reduced mass,
$2 \mu_{np} = 2 M_p M_n / (M_p + M_n)$.
Therefore, for the OBE $NN$ potential we have 
\begin{eqnarray}
V_{np} (r) &=& V_{np}^{1\pi} (r) + V^{1 \sigma} (r) + 
V_{np}^{1\rho} (r) + V^{1 \omega} (r) + \dots  \, ,  \nonumber \\
V_{nn} (r) &=& V_{nn}^{1\pi} (r) + V^{1 \sigma} (r) + 
V_{nn}^{1\rho} (r) + V^{1 \omega} (r) + \dots  \, , \nonumber \\
V_{pp} (r) &=& V_{pp}^{1\pi} (r) + V^{1 \sigma} (r) + 
V_{pp}^{1\rho} (r) + V^{1 \omega} (r) + \dots \, . \nonumber \\
\end{eqnarray} 
Clearly, the potentials in the different channels are not very
different from one to another {\it quantitatively}.
Actually, the $\sigma$ and $\omega$ exchange contributions
coincide identically.
On the other hand the $\pi$ and $\rho$ take into account the different
charged mesons which are exchanged.
Obviously, one expects the symmetry breaking effects coming from
$\pi$ exchange to be more important than $\rho$ exchange. 
Theoretical computations seem to support the previous, giving
$\Delta \alpha_{{\rm CIB}, \pi} = 3.24\,{\rm fm}$ and
$\Delta \alpha_{{\rm CIB}, \rho} = -0.29\,{\rm fm}$, see Ref.~\cite{Li:1998xh}.
As a consequence $\rho$ mass differences are negligible.

The long distance correlation between the scattering length and effective range
looks as 
\begin{eqnarray}
r_{0,np} &=& A_{np} + \frac{B_{np}}{\alpha_{0,np}}+ 
\frac{C_{np}}{\alpha_{0,np}^2} \, , \\ 
r_{0,pp} &=& A_{pp} + \frac{B_{pp}}{\alpha_{0,pp}}+ 
\frac{C_{np}}{\alpha_{0,np}^2} \, , \\ 
r_{0,nn} &=& A_{nn} + \frac{B_{nn}}{\alpha_{0,nn}}+ 
\frac{C_{np}}{\alpha_{0,np}^2} \,   \\ 
r_{0,pp}^C &=& A_{pp}^C + \frac{B_{pp}^C}{\alpha_{0,C,pp}}+ 
\frac{C_{pp}^C}{\alpha_{0,C,pp}^2} \, ,  
\end{eqnarray}
while the phase shifts are given by
\begin{eqnarray}
k \cot \delta_{0,np} &=& \frac{ \alpha_{0,np} {\cal A}_{np} ( k) +
{\cal B}_{np} (k)}{ \alpha_{0,np} {\cal C}_{np} ( k) + {\cal D}_{np} (k)}
\, ,\\
k \cot \delta_{0,nn} &=& \frac{ \alpha_{0,nn} {\cal A}_{nn}
( k) + {\cal B}_{nn} (k)}{ \alpha_{0,nn}
{\cal C}_{nn} ( k) + {\cal D}_{nn} (k)} \, , \\
k \cot \delta_{0,np} &=& \frac{ \alpha_{0,np}
{\cal A}_{np} ( k) + {\cal B}_{pp} (k)}
{ \alpha_{0,pp} {\cal C}_{pp}( k) + {\cal D}_{pp} (k)} \, , \\ 
C^2(\eta)\,k\,\cot{\delta^{C}_{0,pp}} &+& \frac{2}{a_B}\,h(\eta) \nonumber\\
&=& \frac{
  \alpha_{0,pp}^C {\cal A}_{pp}^C ( k) + {\cal B}_{pp}^C (k)}{
  \alpha_{0,pp}^C {\cal C}_{pp}^C ( k) + {\cal D}_{pp}^C (k)} \, .
\label{eq:phase_all_singlets}
\end{eqnarray} 
We remind that the scattering lengths are {\it independent} of the
potentials.

In Fig.~\ref{fig:ABCD} we show the universal functions ${\cal A}$,
${\cal B}$ ${\cal C}$ and ${\cal D}$ for the four cases considered. 
As can be seen, for $nn$, $np$ and $pp$ they coincide even though
the potentials are {\it different}. This means in particular that most
of the CIB and CSB effects for $p \le 400 {\rm MeV}$ come solely from
the difference in the scattering length (there are no genuine sizeable
effective range effects).
It is also interesting to see that the Coulomb corrections to the $pp(c)$
universal functions differ increasingly for higher energies.

\section{The short distance connection}
\label{sec:SDC}

As it is well known, at large energies the $nn$, $np$ and $pp$
phase-shifts start to resemble each other, which means that charge
invariance is respected for large enough momenta.
Most of the charge invariance and charge symmetry breaking effects only
affect the low energy behaviour, specifically the scattering lengths,
where one finds $\Delta \alpha_{\rm CIB} = 5.7(3)\,{\rm fm}$ and
$\Delta \alpha_{\rm CSB} = 1.5(5)\,{\rm fm}$.
When one considers the effective range, the symmetry breaking effects
are already ten times smaller than in the scattering length case,
being of the order of the tenth of a fermi.
The problem is how to explain these differences.

In the traditional approach all the CIB and CSB effects are explained via
the OBE potential, Eq.~(\ref{eq:potential-OBE}).
The Schr\"odinger equation is integrated from the origin to infinity with
regular boundary conditions and all the difference between scattering
observables must come from the potential.
In the renormalization approach things get more involved: there are explicit
contributions coming from short distance operators which are used to weaken
the short distance sensitivity.
The problem is how to implement either charge independence or its breaking
within this approach in a regulator independent way.
If we assume that at lowest order all the charge independence breaking
comes from the finite range potential, one is tempted to identify 
short distance charge independence with identical logarithmic
boundary conditions.
For example, if we relate the $nn$ and $np$ problems with
\begin{eqnarray}
\frac{u_{nn}'(r_c)}{u_{nn}(r_c)} = \frac{u_{np}'(r_c)}{u_{np}(r_c)} \, ,
\end{eqnarray}
we will find that this relation produces log-divergent results for the OBE
potential in the limit $r_c \to 0$.
Another option is to regulate with a short distance delta potential
\begin{eqnarray}
V_C(r; r_c) = \frac{C_0(r_c)}{4 \pi r_c^2}\,\delta(r - r_c) \, ,
\end{eqnarray}
which corresponds to a specific regularization of the $\delta$ function
potential, and assume that charge independence at short distance
is equivalent to $C_{0,nn}(r_c) = C_{0,np}(r_c) = C_{0,pp}(r_c) = C_{0}(r_c)$.
This choice leads to the following logarithmic boundary condition
between $nn$ and $np$
\begin{eqnarray}
\frac{1}{M_{n}}
\left( \frac{u_{nn}'(r_c)}{u_{nn}(r_c)} - \frac{1}{r_c} \right) = 
\frac{1}{M_{np}}
\left( \frac{u_{np}'(r_c)}{u_{np}(r_c)}  - \frac{1}{r_c} \right) \, ,
\end{eqnarray}
where $M_n$ is the neutron mass and $M_{np}$ is twice the reduced $np$ mass.
The counterterm conditions also runs into the same cut-off dependence problems
than the logarithmic boundary condition.
This means in particular that the two previous proposals are regulator
dependent, and hence model dependent, and pose a serious problem on
what is meant by charge independence of short distance operators.
We will show that by using the hypothesis of charge independence 
at short distances together with finiteness,
a relation between them can be established which works rather satisfactorily.

At short distances all the pp (strong/Coulomb), np and nn potentials 
have an attractive Coulomb like behaviour
\begin{eqnarray} 
2\mu_{NN}\,V_{NN}(r) \xrightarrow[r\to 0]{} -\frac{1}{R\,r} \, , 
\end{eqnarray} 
where $NN$ either refers to $pp$ (strong), $pp$ (Coulomb),
$np$ or $nn$, and $\mu_{NN}$ and $V_{NN}$ are the corresponding
reduced mass and potential.
The constant $R$ depends on the problem; for the OBE potential
of Eq.(\ref{eq:potential-OBE}) with the additional simplification of
taking $m_{\omega} = m_{\rho}$ and defining $g_{\omega NN}^*$,
we get the scales 
\begin{eqnarray}
\frac1{R_{np}} &=& M_{np} \left( f_{\pi NN}^2 + g_{\sigma NN}^2 -
{g_{\omega NN}^*}^2\right) \, , \\
\frac{1}{R_{nn}} &=& M_n 
\left( f_{\pi NN}^2 + g_{\sigma NN}^2 -{g_{\omega NN}^*}^2 \right) \, , \\
\frac{1}{R_{pp}} &=&
M_p \left( f_{\pi NN}^2 + g_{\sigma NN}^2 - {g_{\omega NN}^*}^2 \right)
\, , \\ 
\frac1{R_{pp}^C} &=& M_p \left(
f_{\pi NN}^2 + g_{\sigma NN}^2 -{g_{\omega NN}^*}^2-\alpha \right) \, ,
\end{eqnarray}
with $M_{np}$ twice the reduced $np$ mass, $M_{np} = 2\,\mu_{np}$.
As a consequence of the short distance Coulomb singularity,
the wave function at short distances approximately behaves
as linear combinations of {\it attractive} Coulomb wave functions
\begin{eqnarray}
u_{k,NN}(r) &\to& a\,
\frac{R}{2}\,\sqrt{x}\,J_1(2\sqrt{x}) + 
b \,2\,\sqrt{x}\,Y_1(2\sqrt{x}) \nonumber \\ 
&+& {\mathcal O}(m r, m R, k^2 r^2, r / R) \, , \nonumber\\
\end{eqnarray}
where now the Bessel functions $J_1$ and $Y_1$ are used
(instead of $I_1$ and $K_1$).
The constants $a$ and $b$ determine the correct linear combination, 
$R$ can either be $R_{nn}$, $R_{np}$, and $R_{pp}$ (strong/Coulomb),
$x = 2 r / R$ and $m$ generically denotes the mass
of any of the exchanged bosons.
The expected $m R$ contributions will only shift the irregular solutions by a
constant.

The previous behaviour can be quite problematic as we can see if we consider
the log-derivative of the wave function at small enough cut-off radii,
which behaves as
\begin{eqnarray} 
R\,\frac{u_{k,NN}'(r_c)}{u_{k,NN}(r_c)} \to
- 2\gamma - \frac{\pi}{4}\,{R}\,\lambda - \log{\frac{r_c}{R}} + \dots 
\end{eqnarray}
where $\lambda = a_0 / b_0$ and the dots refer to higher order terms,
like $m r_c$ or $k^2 r_c^2$ corrections.
With this behaviour, we can see that naively identifying the log-derivative at
the cut-off radius in order to obtain correlations between observables of
the different two nucleon systems will yield divergent results.
For example, relating $np$ and $nn$
\begin{eqnarray} 
\frac{u_{k,nn}'(r_c)}{u_{k,nn}(r_c)} = \frac{u_{k,np}'(r_c)}{u_{k,np}(r_c)} \,,
\end{eqnarray}
generates the singularity
\begin{eqnarray} 
\frac{1}{R_{np}}\log{\left( \frac{r_c}{R_{np}} \right)} - 
\frac{1}{R_{nn}}\log{\left( \frac{r_c}{R_{nn}} \right)} \,.
\end{eqnarray}
This singularity is indeed mild, as it can only be seen at very short distances
(depending on how small is the difference between $1/R_{nn} - 1/R_{np}$),
but sooner or later will ruin our results.

\begin{figure*}[ttt]
\includegraphics[height=5.5cm,width=5.5cm,angle=270]{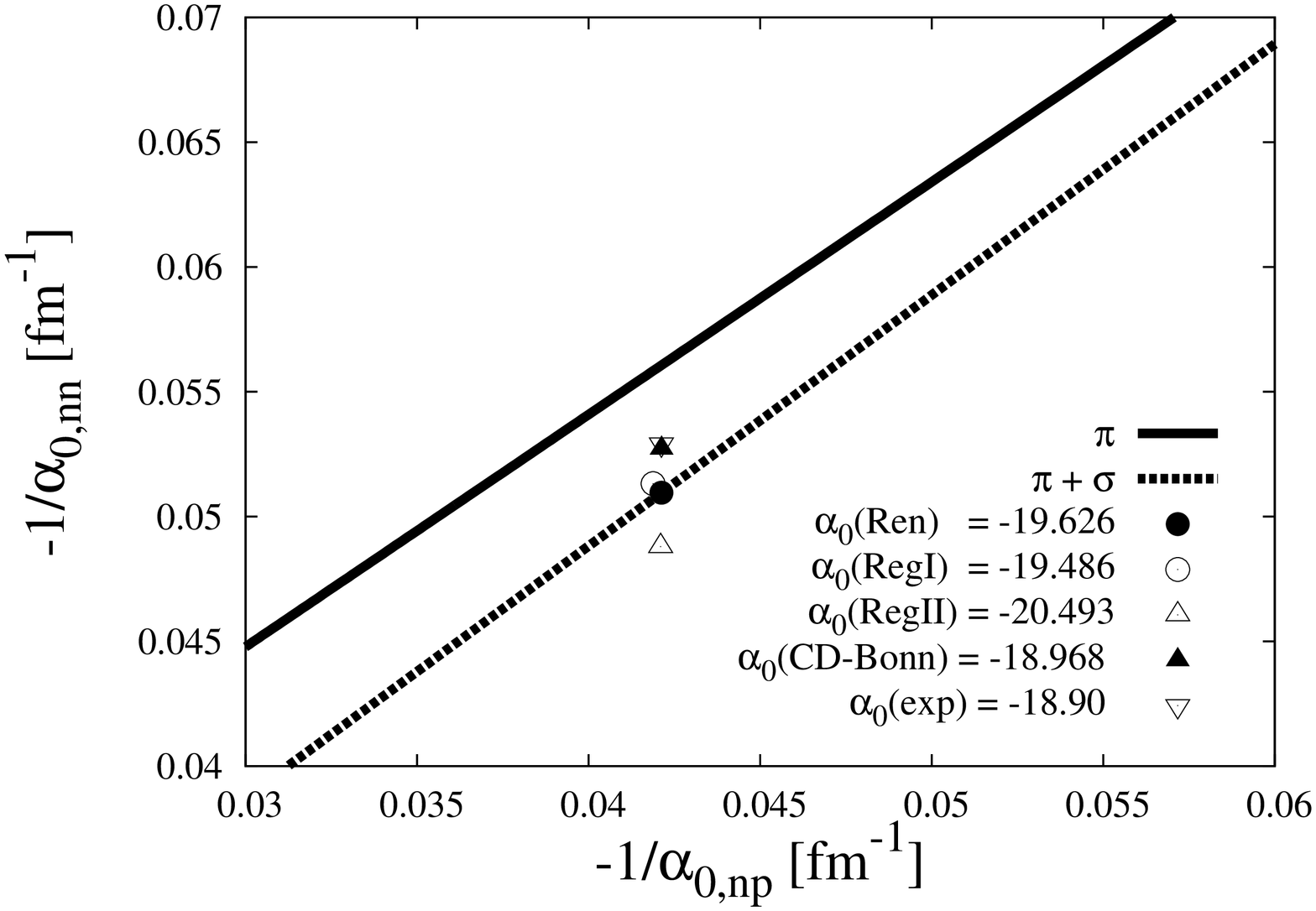}
\includegraphics[height=5.5cm,width=5.5cm,angle=270]{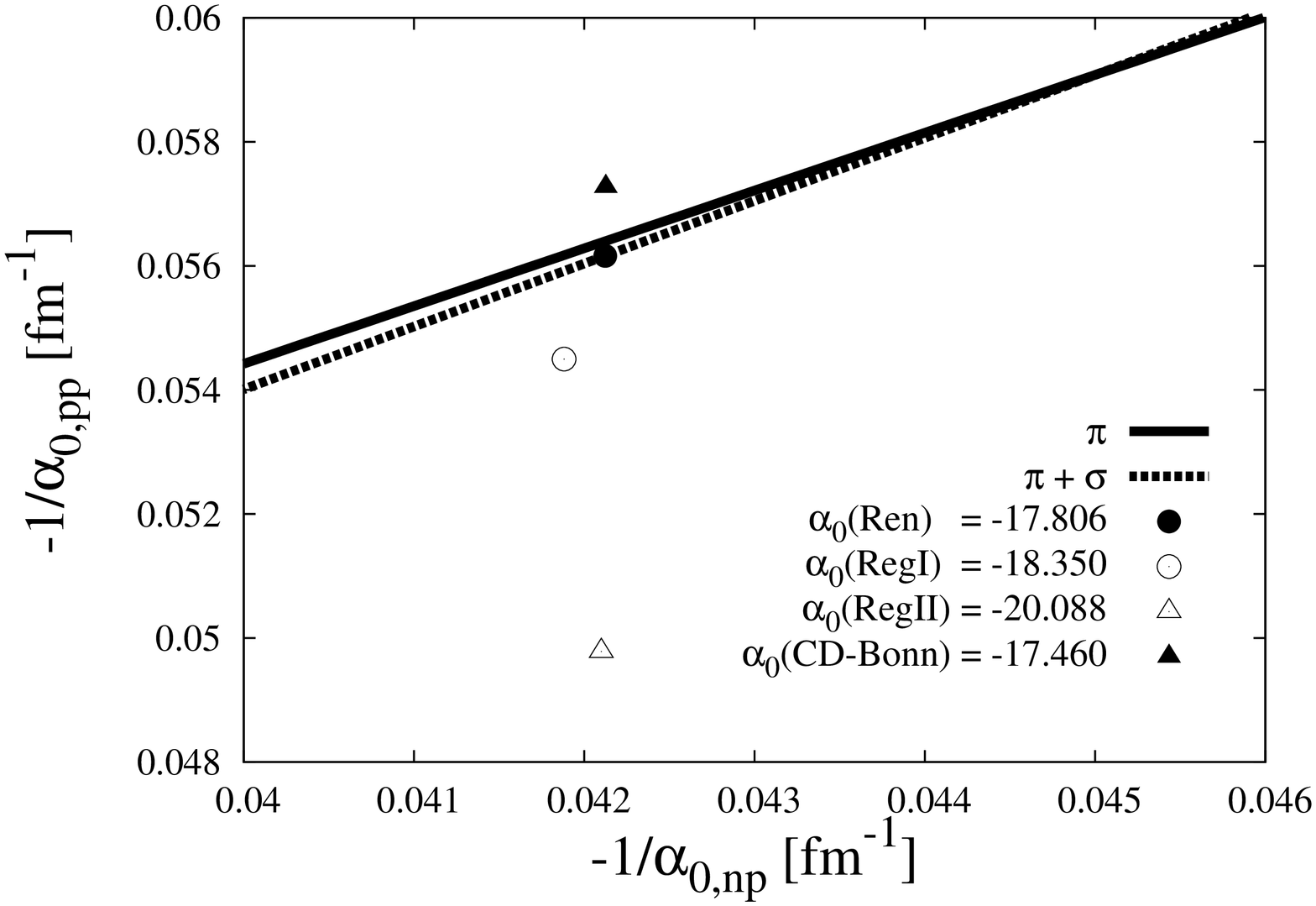} 
\includegraphics[height=5.5cm,width=5.5cm,angle=270]{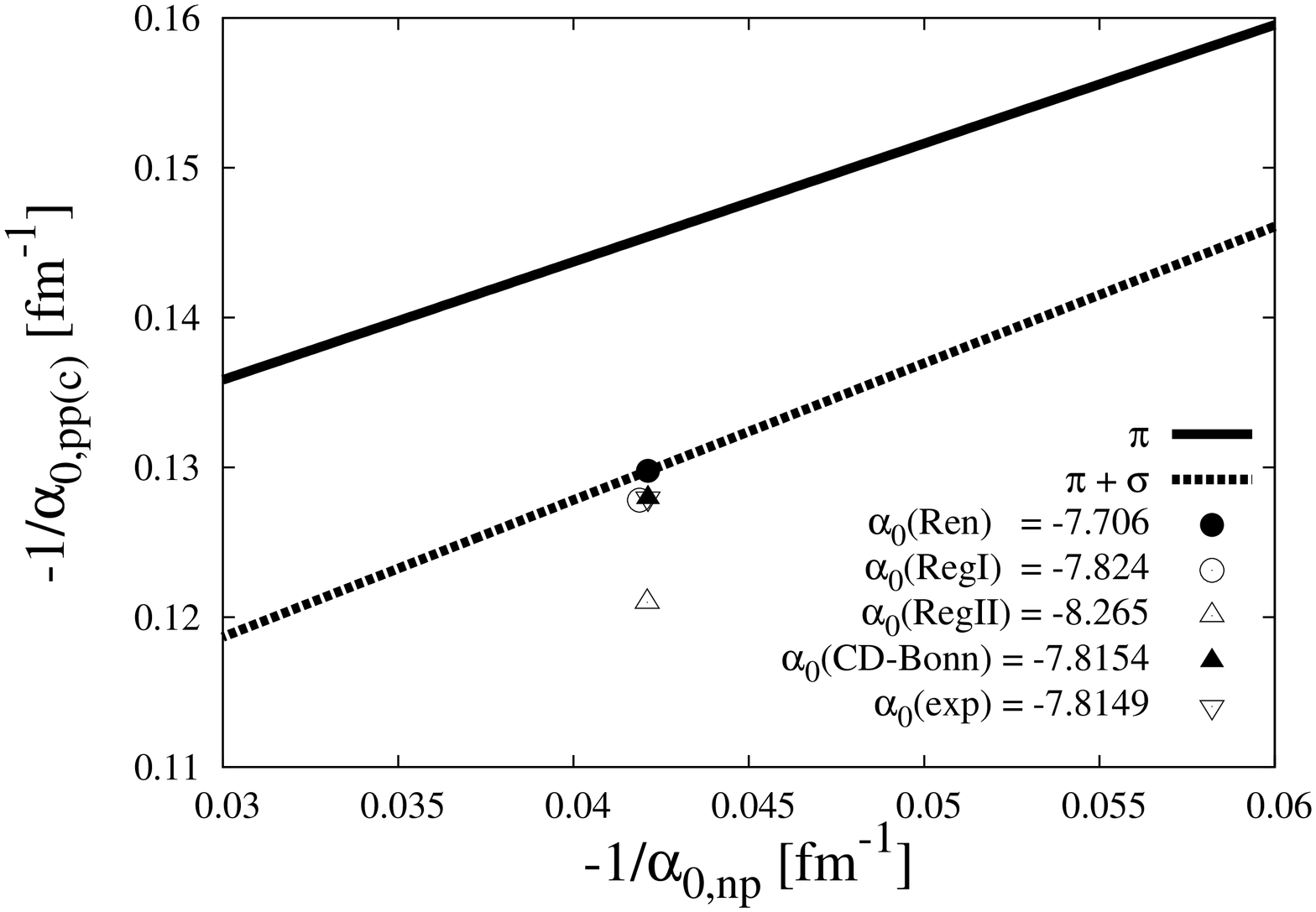}
\caption{The relations between the {\it predicted} scattering lengths
  in the $^1S_0$ channel for $nn$,$pp$,$pp(c)$ as a function of
  $\alpha_{0,np}$ when the successive $\pi$, $\pi+\sigma$ and
  $\pi+\sigma+\rho+\omega$ contributions are included. We plot inverse
  scattering lengths. Note the small scale.}
\label{fig:alphas}
\end{figure*}

Under these circumstances there is a quantity that can be constructed
from the log-derivative at short distance that is finite
in the $r_c \to 0$ limit.
This quantity is the following 
\begin{eqnarray} 
{\cal S} = R \,\frac{u'(r_c)}{u(r_c)} +
\log{\left( \frac{r_c}{R} \right)}
\quad,  \, r_c \ll R \, ,
\end{eqnarray}
which is cut-off and energy independent.
This suggests that different scattering problems,
having different short distances constants but the same logarithmic scale
dependence, can be connected in such a way that the scale dependence
is eliminated.
This is done by equating the corresponding ${\cal S}$'s
\begin{eqnarray}
{\cal S}_{1} = {\cal S}_{2} \, ,
\end{eqnarray}
where ${1}$ and ${2}$ refer to two different
$NN = nn$, $np$, $pp$, $pp(c)$ cases.
We can give here two examples of the adequacy of the short distance connection.
The first one is to obtain the strong $pp$ scattering length from the
experimental Coulomb one, $\alpha^C_{0,pp} = -7.8149\,{\rm fm}$ yielding 
$\alpha_{0,pp} = -18.46\,{\rm fm}$, a not unreasonable results
(to be compared with the extraction $\alpha_{0,pp} = -17.3\,{\rm fm}$,
see Ref.~\cite{Miller:1990iz}, where the error comes from model-dependence).
The CD-Bonn potential gives a value of $\alpha_{0,pp} = -17.46\,{\rm fm}$.
The extracted effective ranges are  $r^C_{0,pp} = 2.735\,{\rm fm}$ and
$r_{0,pp} = 2.789\,{\rm fm}$. 
As a second example, by taking the $np$ scattering length as input,
$\alpha_{0,np} = -23.74\,{\rm fm}$, we can obtain all
the $NN$ scattering lengths, giving $\alpha_{0,nn} = -19.626\,{\rm fm}$,
$\alpha_{0,pp} = -17.806\,{\rm fm}$ and $\alpha^C_{0,pp} = -7.706\,{\rm fm}$ 
for the scattering lengths and $r_{0,np} = 2.672 \,{\rm fm}$, $r_{0,nn} = 2.771
\,{\rm fm}$, $r_{0,pp} = 2.802 \,{\rm fm}$ and  $r^C_{0,pp} = 2.747  \,{\rm
fm}$ for the effective ranges.
A remarkable aspect of the previous computation is that one obtains $
\Delta \alpha_{\rm CIB} = 5.024\,{\rm fm} $, $ \Delta r_{\rm CIB} =
0.115\,{\rm fm} $, $\Delta \alpha_{\rm CSB} = 1.82\,{\rm fm} $ and $
\Delta r_{\rm CSB} = 0.031\,{\rm fm} $ which agree within error
estimations with the expected values for these two
quantities~\cite{Miller:1990iz}. In Table~\ref{tab:low_par} we
summarize the results obtained with the short distance connection
(renormalized) and the one obtained integrating upward with a regular
boundary condition (regular). We can see that in the case of a big
$g_{\omega NN}^*$ the regular solution does a poor job in calculation
the low energy parameters (LEP) in other channels. The CD-Bonn
potential~\cite{Machleidt:2000ge} corresponds with this scenario,
i.e., a big SU(3) breaking coupling constant but with any spurious
bound state.  Looking at this table one can understand why in this
model a different mass for a ficticious $\sigma$-meson is used in each
$NN$ channel. The strong fine-tuning that appears in this situation
hinders the relation between different $NN$ problems.  

A further interesting example of the adequacy of the short distane
connection is illustrated in Appendix~\ref{sec:ppF} where the
Gamow-Teller matrix element appearing in the proton-proton fusion
process is analyzed.

Note, however, that the previous is not the only possible {\it covariant}
short distance connection, as we could have defined
\begin{eqnarray} 
{\cal S}' = R \,\frac{u'(r_c)}{u(r_c)} +
\log{\left( \frac{\lambda\,r_c}{R} \right)}
\quad,  \, r_c \ll R \, ,
\end{eqnarray}
with $\lambda$ some arbitrary constant, which depends on the specific $NN$
problem which is being considered.
A natural choice is to take $\lambda$ of order unity, 
which does not make much difference between different choices of ${\cal S}$
due to the weak logarithmic behaviour.
It must be stressed though that the results are not unique:
arbitrary $\lambda$'s can be introduced to better connect 
the different two nucleon systems.
As the hypothesis of charge dependence of short distance operator cannot be
implemented in a completely model independent way, we will chose to take
$\lambda_{nn} = \lambda_{np} = \lambda_{pp}$ at first order.
We have already seen
that this simple condition generates quite accurate results,
meaning that corrections due to $\Delta \lambda$ are indeed small, and
can be effectively considered as higher order effects, confirming thus
our expectations.
As an example of what values of $\Delta \lambda$ to expect, if we try to
correlate the strong and Coulomb scattering lengths, we will get
$\lambda_{pp} - \lambda_{pp}^C = 0.0321-0.0471$, where
the range given is a consequence of the uncertainty in
$\alpha_{pp} = -17.3(4)\,{\rm fm}$.

To clarify the implications of the short distance connection, let us
consider two different problems $1$ and $2$, which have the associated
Coulomb length scales $R_1$ and $R_2$.
In other words, we have the differential equations
\begin{eqnarray}
-u_1'' + 2\mu_1\,V_1(r) u_1(r) &=& k^2 u_1(r) \, , \\
-u_2'' + 2\mu_2\,V_2(r) u_2(r) &=& k^2 u_2(r) \, ,
\end{eqnarray}
where the reduced potentials behave as $1/r$ at short distances
\begin{eqnarray}
2\mu_1\,V_1(r) &\to& - \frac{1}{R_1\,r} \, , \\
2\mu_2\,V_2(r) &\to& - \frac{1}{R_2\,r} \, .
\end{eqnarray}

\begin{centering}
\begin{table*}[ttt]
\begin{tabular}{|c|c|c|c|c|c|c|}\hline\hline
{$NN$} & LEP &
Renor. &
Reg. -I  &
Reg. -II &
CD-Bonn~\cite{Machleidt:2000ge} &
Exp.~\cite{Machleidt:2000ge} \\ 
 &  & ($g_{\omega NN}^* \sim 0$) & ($g_{\omega NN}^* \sim 8$)
 & ($g_{\omega NN}^*\sim 20$) & & \\
\hline
$np$
 & $\alpha_0$ [fm] & {\rm input} & -23.737 & -23.738 & -23.738 & -23.74(2)\\
 & $r_0$ [fm]      & 2.672       &   2.678 &   2.677 &   2.671 &   2.77(5)\\
\hline  
$pp$
 & $\alpha_0 $ [fm] & -17.806 & -18.350 & -20.088 & -17.46  & $-$\\
 & $r_0 $      [fm] &   2.802 &   2.799 &   2.768 &   2.845 & $-$\\
\hline  
$pp(c)$
 & $\alpha_0$ [fm]& -7.706 & -7.824 & -8.265 & -7.8154 & -7.8149(29)\\
 & $r_0$ [fm]     &  2.747 &  2.641 &  2.693 &  2.773  & 2.769(14)  \\
\hline  
$nn$
 & $\alpha_0$ [fm] & -19.626 & -19.486 & -20.493 & -18.968 & -18.9(4)\\ 
 & $r_0 $ [fm]     &   2.771 &   2.780 &   2.763 &   2.819 & 2.75(11)\\
\hline \hline
\end{tabular}
\caption{\label{tab:low_par} $NN$ low-energy parameters in the different
  scenarios. Renormalization only needs the $np$ scattering length as an
  input parameter, all the other are calculated without ambiguities. The OBEP
parameters have been fitted in the $np$ case and kept the same in the other
cases. Here (c) means Coulomb interaction is switch on.}
\end{table*}
\end{centering}

\begin{figure*}[ttt]
\begin{center}
\epsfig{figure=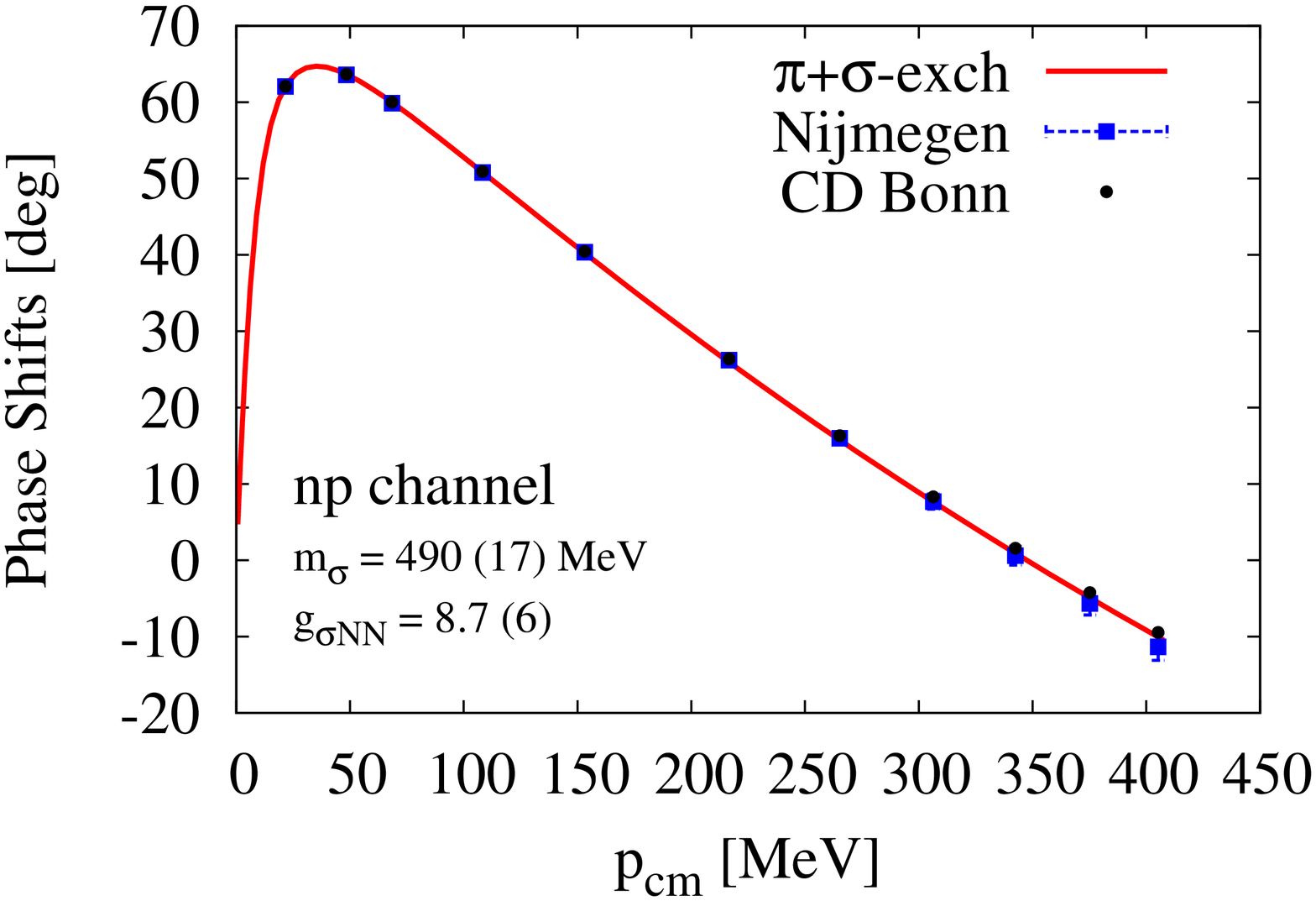,height=5.5cm,width=6cm,angle=270} 
\epsfig{figure=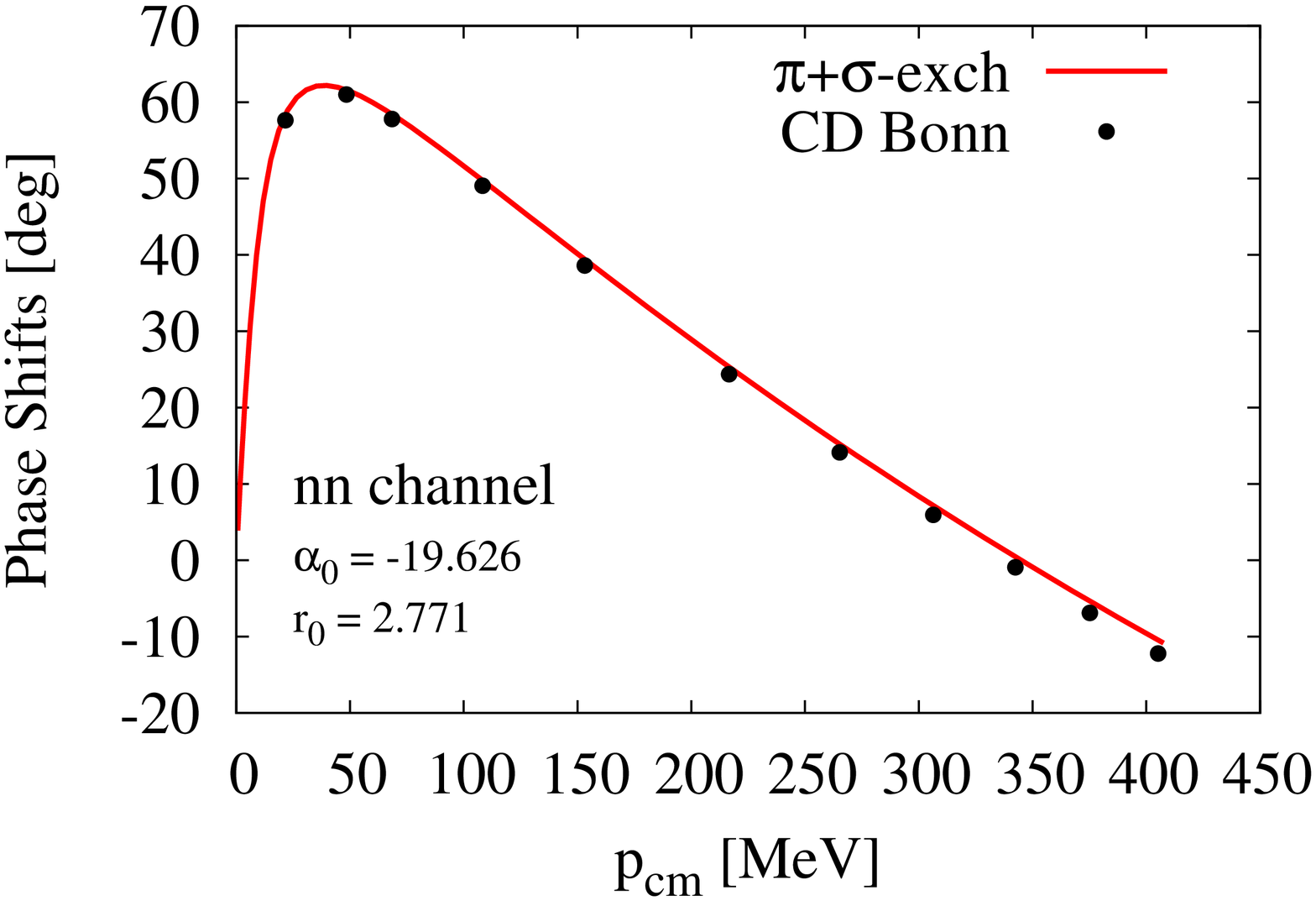,height=5.5cm,width=6cm,angle=270}
\epsfig{figure=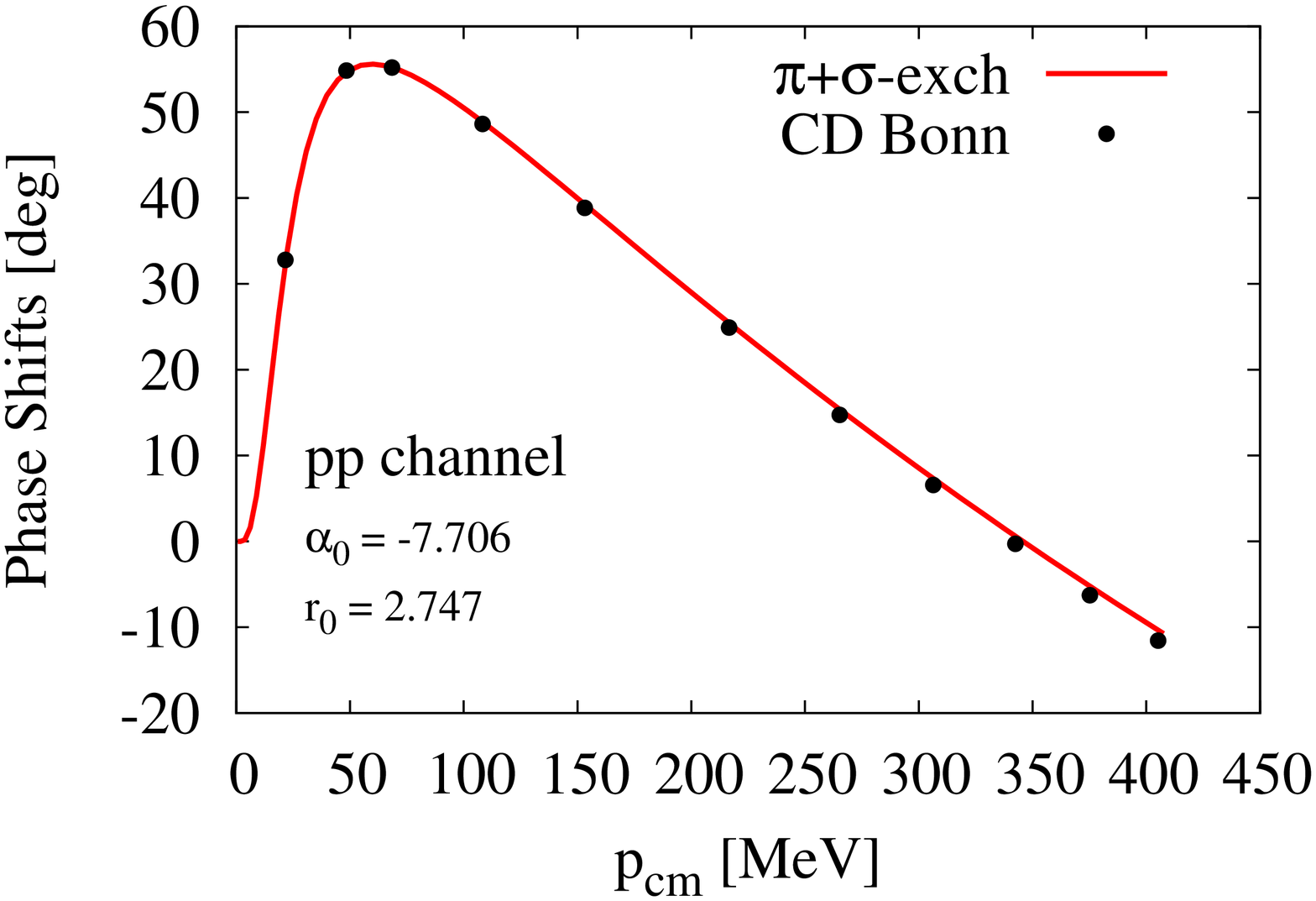,height=5.5cm,width=6cm,angle=270}
\end{center}
\caption{Renormalized phase shifts for the OBE potential with CSB OPE
  + $\sigma$ as a function of the c.m. momentum in the singlet $^1S_0$
  channel. In the left panel we show the fitted $np$ phase-shift to
  the Nijmegen results~\cite{Stoks:1994wp}. In the middle and right
  panels the {\it predicted} $pp(c)$ and $nn$ are depicted and
  compared to the CD-Bonn result~\cite{Machleidt:2000ge}.}
\label{fig:phase-shift_1S0}
\end{figure*}

These two problems are related at short distances through
the boundary condition corresponding
to the short distance connection ${\cal S}_1 = {\cal S}_2$
\begin{eqnarray} 
\label{eq:connection-explicit}
 R_2\,\frac{u'_2(r_c)}{u_2(r_c)} =
\log{\frac{R_1}{R_2}} + R_1 \frac{u'_1(r_c)}{u_1(r_c)} \, .
\end{eqnarray}
If we have only fixed the scattering length, the above condition becomes
energy independent when the cut-off is small enough, which means that it
can be evaluated with the zero-energy wave functions of the two-body
systems $1$ and $2$.
Using the superposition principle, the previous zero energy wave functions
can be written as
\begin{eqnarray}
u_1(r) = v_1(r) - \frac{1}{\alpha_1} w_1(r) \, ,\\
u_2(r) = v_2(r) - \frac{1}{\alpha_2} w_2(r) \, .
\end{eqnarray}
These wave functions can be included in Eq.~(\ref{eq:connection-explicit}),
yielding the following relation between the scattering lengths $\alpha_1$ and
$\alpha_2$ of the {\it two different} problems
\begin{eqnarray}
\frac{A}{\alpha_1} = \frac{B}{\alpha_2} + C + \frac{D}{\alpha_1\,\alpha_2} \, .
\end{eqnarray}
Therefore, if we make the hypothesis of Charge independence at short
distances~\footnote{Generally one might expect  
$$
{\cal S}_{pp}^{C} = {\cal S}_{pp}^{S} + \alpha {\cal S}_{pp}^{(1)} + \dots  
$$. Our results are consistent with the expected smallness of the
corrections.}
\begin{eqnarray}
{\cal S}_{np} = {\cal S}_{nn} = {\cal S}_{pp} = {\cal S}_{pp}^C \, ,
\label{eq:Cs-short-distance}
\end{eqnarray}
and by making use of the superposition principle, we can write 
\begin{eqnarray}
u_{0,np} (r) &=& v_{0,np} (r) - \frac{1}{\alpha_{0,np}} w_{0,np} (r) \, , \\ 
u_{0,nn} (r) &=& v_{0,nn} (r) - \frac{1}{\alpha_{0,nn}} w_{0,nn} (r) \, , \\ 
u_{0,pp} (r) &=&
v_{0,pp} (r) - \frac{1}{\alpha_{0,pp}} w_{0,pp} (r) \, , \\ 
u_{0,pp}^C (r) &=&
v_{0,pp}^C (r) - \frac{1}{\alpha_{0,pp}^C} w_{0,pp}^C (r) \, , 
\end{eqnarray}
so we get the bilinear relations between all scattering lengths
\begin{eqnarray}
\frac{A_{nn}}{\alpha_{nn}} &=& \frac{B_{nn}}{\alpha_{np}} +
C_{nn} + \frac{D_{nn}}{\alpha_{nn}\,\alpha_{np}} \, , \\
\frac{A_{pp}}{\alpha_{pp}} &=&  \frac{B_{pp}}{\alpha_{np}} + 
C_{pp} + \frac{D_{pp}}{\alpha_{pp}\,\alpha_{np}} \, , \\
\frac{A^C_{pp}}{\alpha^C_{pp}} &=&  \frac{B^C_{pp}}{\alpha_{np}} + 
C^C_{pp} + \frac{D^C_{pp}}{\alpha^C_{pp}\,\alpha_{np}} \, ,
\end{eqnarray}
etc. In Fig.~\ref{fig:alphas} we show the dependence of the scattering
lengths as obtained from the $np$ scattering length and the previous
correlations.
As can be seen, the correlations work rather well, confirming the idea
that finiteness is a good criterion to implement charge independence
of short distance operators. Numerical values are listed in
Table~\ref{tab:low_par} for the experimental value of $\alpha_{0,np}$.

It is interesting to see how the short distance connection works at
finite energy and, in particular, if a given specific $NN$ channel is
able to predict the phase shifts for the remaining channels. In
Fig.~\ref{fig:phase-shift_1S0} we plot the extracted $nn$ and $pp(c)$
phase shifts when the OBE parameters have been fixed from the $^1S_0$
Nijmegen $np$ phase shifts. We have computed these phase shifts
renormalizing in the $np$ channel, i.e., fixing $\alpha_{0,np}$ as
input and integrating inward the Schr\"odinger equation, and then
using Eq.~\eqref{eq:Cs-short-distance} we connect with the other
channels.
As we can see the short distance connection can be used to predict the $^1S_0$
phase shits for the rest of the channel with a high degree of accuracy.

\section{Conclusions} 
\label{sec:concl}

In this paper we have analyzed the charge dependence and charge
symmetry breaking of the $NN$ interaction. We have used the OBE model
with exchange of $\pi$, $\sigma$, $\omega$ and $\rho$ mesons and we
have implemented CSB by means of pion mass splitting in the OPE
potential and different nucleon masses.  In particular, and as in
previous works, we have selected the $^1S_0$ $np$ channel to fit
scalar meson parameters, $m_\sigma$ and $g_{\sigma NN}$, as well as
vector meson couplings, $g_{\omega NN}$ and $f_{\rho NN}$ to the
Nijmegen phase shifts~\cite{Stoks:1994wp}. A fine tuning problem
arises when we using the customary regular boundary condition at the
origin $u(0) = 0$. This problem appears in all $np$,$nn$,$pp$ and
$pp(c)$ channels and large,$\sim 40\%$ , violations of $SU(3)$ values
of the $g_{\omega NN}$ coupling constant are needed. Traditionally a
great amount of effects such as multi-meson exchanges have been
essential to explain the differences in phase shifts and threshold
parameters for all $np$,$nn$,$pp$ and $pp(c)$
channels~\cite{Li:1998ya,Li:1998xh,Machleidt:2000ge} or the role
played by $\rho-\omega$~\cite{McNamee:1974vb,Coon:1977gj,Friar:1977is}
and/or $\pi-\eta$~\cite{Piekarewicz:1993ad} mixing were invoked. These
standard approaches need very precise information on the interaction
at all distances.

However, once we admit incomplete knowledge of the interaction at
short distances, it is possible to sidestep the problem of fine tuning
by imposing a renormalization condition; at any stage of the
calculation the scattering length is always kept fixed. This
renormalization approach embodies short distance insensitivity.  As a
consequence, in the Charge Independent case, one can confortably take
the experimental and/or SU(3) values for vector meson couplings.  For
the same reason we can only hope to quantitatively describe the
relative changes due to the Charge Symmetry Breaking of the
interaction at long distances. These considerations alone allow to
extract some universal information on the symmetry breaking pattern
where the $np$, $nn$ and $pp$ channels look very much the same at all
energies even though the potentials are different and are indeed CSB.
We have used a short distance condition to relate the renormalized
$np$ channel with the others $nn$, $pp$ and $pp(c)$. This short
distance connection is so far an assumption based on finiteness but we
have seen that reasonable results are obtained for low energy
parameters and phase shifts. Our predictions for $(\Delta
\alpha_{CIB},\ \Delta r_{CIB})$ and $(\Delta \alpha_{CSB},\ \Delta
r_{CSB})$ are compatible with the empirical one within the error
estimation. This is in fact a remarkable result: all channels are
generated with {\it just one} scattering length, say $np$, and the
long distance components of the potential where the CSB is, via
physical pion and nucleon masses, explicitly built in.  Our result is
compatible with the interpretation that (relative) CSB sits at large
distances. The absolute CSB is in a sense as uncertain as the short
distance components of the NN force itself and cannot be determined
independently of the Charge Invariant interaction.

\vskip1cm
\begin{acknowledgments}
We thank J. Haidenbauer for a critical reading of the ms and
L.L. Salcedo for providing his fortran code on Coulomb wave functions.
This work is supported Supported by the Spanish DGI and FEDER funds
with grant FIS2008-01143/FIS, Junta de Andaluc{\'\i}a grant FQM225-05,
Spanish Ingenio-Consolider 2010 Program CPAN (CSD2007-00042) and by
the EU Research Infrastructure Integrating Initiative HadronPhysics2.
\end{acknowledgments}

\appendix 

\section{Proton-proton fusion}
\label{sec:ppF}

We would like to analyze further consequences of the short distance
connection assumed by Eq.~\eqref{eq:Cs-short-distance}. An interesting
process is the proton-proton fusion reaction $pp \to d\ e^+ \nu_e$
which is of central importance to stellar physics and neutrino
astro-physics. In fact, it is the dominant solar neutrino source. The
temperature in the Sun core is around $T_c = 15\times 10^6 K$ which
means that we have protons of momentum $p\sim (2m_p T_c)^{1/2} \sim
1.1\, {\rm MeV}$. At these low energies, the reaction is dominated by
the $^1S_0\to d$ nuclear transition. The Gamow-Teller (GT) matrix
element for this process (without MECs) is given by,
\begin{eqnarray}
 A_S M_{GT} = \int_0^{\infty} {\rm dr}\ u_\gamma(r) u_{0,pp}(r)
\end{eqnarray}
where $u_{0,pp}$ is the zero energy reduced wave function for the $pp(c)$
system which can be related with the $np$ problem by
Eq.~\eqref{eq:Cs-short-distance}. Then taking $\alpha_{0,np}$ as input and
integrating in we can calculate $u_{0,pp}$. For deuteron we take as a first
approximation the normalized bound state, 
\begin{eqnarray}
 u_\gamma (r) \to A_S e^{-\gamma_d r}\, , 
\end{eqnarray}
with $\gamma_d=0.2316 {\rm fm}^{-1}$ and integrate inward the Schr\"odinger
equation with negative energy $E=-\gamma_d^2/M_{np}$. We obtain
a value $M_{GT} = 5.189\,{\rm fm}$ to be compared to a more sophisticated
one~\cite{Park:1998wq} using Argonne $V18$ wave functions $M_{GT}|_{AV18} =
4.859\,{\rm fm}$.

\begin{figure}[tbc]
\begin{center}
\includegraphics[height=6.5cm,width=6.5cm,angle=270]{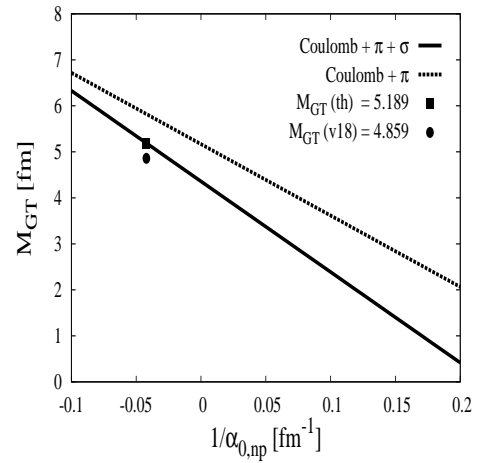}
\end{center}
\caption{Dependence of the $pp$ fusion Gamow-Teller matrix element (in
  ${\rm fm}$) depending on the singlet $np$ inverse scattering length
  $1/\alpha_0$ (in ${\rm fm}^{-1}$) using the short distance connection
Eq.~\eqref{eq:Cs-short-distance}.}
\label{fig:MG-a0}
\end{figure}

In Fig.~\ref{fig:MG-a0} we show the GT matrix element correlation with
the $np$ scattering length compared with the AV18 calculation. Of course we
have not included the tensor force which mixed $S$ and $D$ waves in the
calculation of the deuteron. However we can appreciate that our numbers are not
very far from much more elaborate calculations~\cite{Park:1998wq}.


\end{document}